\documentclass[fleqn,usenatbib]{mnras}

\usepackage{newtxtext,newtxmath}

\usepackage[T1]{fontenc}

\DeclareRobustCommand{\VAN}[3]{#2}
\let\VANthebibliography\thebibliography
\def\thebibliography{\DeclareRobustCommand{\VAN}[3]{##3}\VANthebibliography}

\usepackage{graphicx}	
\usepackage{amsmath}	

\renewcommand{\vec}[1]{\boldsymbol{#1}}

\newcommand\js{\bgroup\markoverwith{\textcolor[rgb]{.0, 0.8, .6}{\rule[0.5ex]{8pt}{1.5pt}}}\ULon}

\newcommand\ls{\bgroup\markoverwith{\textcolor[rgb]{.8, 0.0, .6}{\rule[0.5ex]{8pt}{1.5pt}}}\ULon}

\title[Stability with Friction]{Dynamical Instability in Multi-Orbiter Systems with Gas Friction}

\author[Li, Rodet and Lai]{
Jiaru Li,$^{1,2}$\thanks{E-mail: jiaru\textunderscore li@astro.cornell.edu}
Laetitia Rodet$^{1}$
and Dong Lai$^{1}$
\\
$^{1}$Center for Astrophysics and Planetary Science, Department of Astronomy, Cornell University, Ithaca, NY 14853, USA\\
$^{2}$ Theoretical Division, Los Alamos National Laboratory, Los Alamos, NM 87545, USA
}

\date{Accepted XXX. Received YYY; in original form ZZZ}

\pubyear{2021}

\begin{document}
\label{firstpage}
\pagerange{\pageref{firstpage}--\pageref{lastpage}}
\maketitle

\begin{abstract}
Closely-packed multi-planet systems are known to experience dynamical instability if the spacings between the planets are too small. 
Such instability can be tempered by the frictional forces acting on the planets from gaseous discs. 
A similar situation applies to stellar-mass black holes embedded in AGN discs around supermassive black holes.
In this paper, we use $N$-body integrations to evaluate how the frictional damping of orbital eccentricity affects the growth of dynamical instability for a wide range of planetary spacing and planet-to-star mass ratios.  
We find that the stability of a system depends on the damping timescale $\tau$ relative to the zero-friction instability growth timescale $t_{\rm inst}$.  In a two-planet system, the frictional damping can stabilise the dynamical evolution if $t_{\rm inst}\gtrsim\tau$.  
With three planets, $t_{\rm inst} \gtrsim 10\tau - 100\tau$ is needed for stabilisation.  
When the separations between the planetary orbits are sufficiently small, $t_{\rm inst}$ can be less than the synodic period between the planets, which makes frictional stabilisation unlikely to occur.  
As the orbital spacing increases, the instability timescale tends to grow exponentially on average, but it can vary by a few orders of magnitude depending on the initial orbital phases of the planets. 
In general, the stable region (at large orbital spacings) and unstable region (at small orbital spacings) are separated by a transition zone, in which the (in)stability of the system is not guaranteed.
We also devise a linear map to analyse the dynamical instability of the ``planet + test-mass'' system, and we find qualitatively similar results to the $N$-body simulations.
\end{abstract}




\section{Introduction}  
\label{sec:intro}

Planetary systems with two or more planets can be dynamically unstable if the spacing between the planetary orbits is small \citep[e.g.,][]{Wisdom1980,Gladman1993,Chambers1996,Zhou2007,Smith2009,Funk2010,Deck2013,Pu2015}.
This instability typically results in collisions or strong scatterings between the planets, which have significant impacts on the architecture of the planetary systems \citep[e.g.,][]{Rasio1996,Weidenschilling1996,Lin1997,Ford2001,Adams2003,Nagasawa2011,Petrovich2014,Frelikh2019,Anderson2020,Li2020,Li2021}.  
For example, extrasolar gas giants (``cold Jupiters'') are known to have a broad eccentricity distribution, indicating that these systems have gone through a phase of dynamical instability in their evolution histories \citep[e.g.,][]{Chatterjee2008,Ford2008,Juric2008,Morbidelli2018,Anderson2020}.
The tightly-packed multi-planet systems of super-Earths and mini-Neptunes discovered by the Kepler spacecraft \citep[e.g.,][]{Lissauer2011a,Lissauer2011b,Fabrycky2014,Campante2015} are found to be close to their instability limit \citep{Pu2015,Volk2015}.
This again suggests that the super-Earth systems may have experienced a dynamically active phase, although the long-term evolution of these systems remains an open question \citep[e.g.,][]{Migaszewski2012,Mahajan2014,Tamayo2017,Ormel2017,Obertas2017,Volk2020}.
In the Solar system, it has long been recognised that the current orbital architecture of giant planets may be the results of an early dynamical evolution \citep[e.g.,][]{Tsiganis2005,Liu2022}.

A widely-used criterion to evaluate the stability of a compact planetary architecture is to compare the distance between neighbouring planets to the mutual Hill radius, which scales with $\mu^{1/3}$, where $\mu$ is the mass ratio between the planets and the central star.
For two-planet systems with initially circular orbits, a ``Hill instability'' criterion for the critical planetary semi-major axis separation can be derived using the conservation of the Jacobi constant and a shearing box approach of the trajectories near the conjunctions \citep{Henon1986,Gladman1993}. 
On the other hand, an alternative approach focusing on the growth of orbital chaos due to resonance overlap leads to a critical separation proportional to $\mu^{2/7}$ \citep{Wisdom1980,Duncan1989}.  
The exact value of the exponent is still under debate even for the restricted three-body problem (i.e., one of the planets has a negligible mass), due to the chaotic nature of the system and the blurriness of the stability-instability divide \citep{Petrovich2015,Petit2020}.

For a system comprising more than two planets, stability is never guaranteed, but the timescale for instability to arise depends on the mutual distance between the planets \citep[e.g.,][]{Chambers1996,Smith2009,Lissauer2021}. 
Mean-motion resonances further prevents a clear determination of the stability boundary of multi-planet systems, leading to the recent development of alternative approaches, such as the use of machine-learning algorithms as a predictive tool \citep{Tamayo2020}. 

Closely-packed, unstable multi-planet systems can be a natural product of planet formation and migration in protoplanetary discs. Planet-disc interaction typically damps the planet's eccentricity, thus prevents orbital crossings between planets and suppresses the dynamical instability. 
Several previous works have explored the interactions between planets embedded in gaseous discs in various scenarios of planet growth and migration, showing that a variety of planetary architecture can be produced \citep[e.g.][]{Lee2009,Marzari2010,Matsumura2010,Lega2013,Liu2022}.
However, a quantitative assessment of the effects of gaseous discs on the dynamical instability of multi-planet systems is lacking. 
The main goal of this paper is to systematically evaluate how the strength of planet eccentricity damping due to the disc affects the onset of the dynamical instability. 
To this end, we apply
parameterized frictional forces acting on the planets, and use $N$-body simulations to quantify the onset and growth of instability as a function of the spacing between planets.

Although the problem studied in this paper pertains to planetary
systems, it is also important for understanding the evolution of
stellar-mass black holes (sBHs) embedded in AGN discs \citep[e.g.][]{McKernan2012,Bartos2017,Stone2017,Secunda2019,Tagawa2020,Li2022}.  
In particular, AGN discs can help bringing sBHs circulating around a supermassive BH into close orbits due to the differential migrations of the BHs and migration traps \citep{Bellovary2016}. 
Close encounters between such tightly-packed sBHs may lead to the formation and merger of binary BHs (\citealt{Li2022}; see also \citealt{Secunda2019,Tagawa2020}).
The effects of the AGN disc on the formation and evolution of the binary BHs are uncertain, and generally require hydrodynamical simulations for proper understanding \citep[see][]{Baruteau2011, LiYP2021,LiYP2022,LiRX2022,Dempsey2022}. 
In \cite{Li2022}, we incorporated parameterised weak frictional forces (with the eccentricity damping time $\gtrsim 10^5$ times the orbital period) in the long-term $N$-body integrations of multi-BH systems around a
supermassive BH, and we found that these forces did not lead to enhanced formation of BH binaries. 
However, we did not explore how the initial instability growth is affected for a wide range of damping timescales.

In this paper, to systematically study the onset and growth of instability in multi-orbiter (planets or BHs) systems around a central massive body (star or supermassive BH) with and without frictional forces, we consider a wide range of ``planet'' to ``star'' mass ratios\footnote{In the remainder of this paper, we shall use the terms ``planet'' and ``star'', although they could be ``BH'' and ``supermassive BH''.}, from $\mu=10^{-7}$ to $10^{-3}$. 
As noted above, because of the effect of mean-motion resonances, the stability property of multi-orbit systems does not just depend on the initial orbital spacings in units of the Hill radius ($\propto\mu^{1/3}$).

The rest of this paper is organised as follows.
In Section~\ref{sec:Hill-wRH}, we study the restricted three-body problem (star, planet and test particle), and characterise the defining features of the instability as a function of $\mu$, with and without damping force. 
In Section~\ref{sec:map}, we present an analytical model (algebraic map) based on the shearing-box approximation, which provides insights to the numerical results. 
In Section~\ref{sec:NP}, we examine the multi-planet case and its relation to the restricted problem. 
We summarise our findings in Section~\ref{sec:summary}.

\section{Instability of Restricted Three-Body Problem with Friction}
\label{sec:Hill-wRH}

In this section, we consider a co-planar system of a central star with mass $M$, an inner planet with $m_1 = \mu M$ on a circular orbit ($e_1=0$), and an outer test particle with $m_2 = 0$.
The test particle may experience a frictional force that tends to damp its eccentricity.
Using $N$-body integration, we numerically determine the stability boundary of such a system for various values of $\mu$ and the damping time.

\subsection{Setup of the simulations}
\label{sec:Hill-wRH-setup}

In our numerical simulations, the initial orbital separation between the planet and the test mass is set as
\begin{equation}
a_2 - a_1 = KR_{\rm H},
\end{equation}
where 
\begin{equation}
R_{\rm H} \equiv \frac{a_1+a_2}{2} \left(\frac{m_1+m_2}{M}\right)^{1/3}
\end{equation}
is the mutual Hill radius, and $K$ is a dimensionless constant. 
We henceforth use the initial orbital period of the planet, $P_1$, as the unit of time. 
We consider different combinations of $\mu$ and $K$ and carry out 500 runs for each combination. 
The initial eccentricity of the test particle is $e_2=10^{-5}$, and the initial values of the argument of the periapsis, the longitude of the ascending node, and the mean anomaly are sampled randomly from the range $[0,2\pi]$, assuming they all have uniform distributions.

The systems are simulated using $N$-body software \textsc{REBOUND} \citep{Rein2012} and the \textsc{IAS15} integrator \citep{Rein2015}. 
For each run, when the orbits of $m_1$ and $m_2$ overlap within their mutual Hill radius, i.e. when
\begin{equation}
\label{eq:Hill-wRH-stop-condition}
a_2(1-e_2)-a_1(1+e_1)<R_{\rm H}
\end{equation}
is satisfied by their real-time orbital elements, this system is considered unstable. 
If such instability is found, we stop the simulation immediately and register the time as $t_{\rm inst}$.
Otherwise, the simulation ends when it reaches $t=10^{5}P_1$ and we consider the system ``stable''.

\subsection{Systems with no friction}
\label{sec:Hill-wRH-nf}

We first study the no-friction situation.
With $\mu=10^{-7}$, $10^{-6}$, $10^{-5}$, $10^{-4}$, $10^{-3}$ and initial $K=2.0$, $2.1$, $2.2$, $...$, $4.0$, the whole experiment contains 105 suites of simulations (and 500 runs with randomised initial angles in each suite).

\begin{figure}
    \includegraphics[width=\columnwidth]{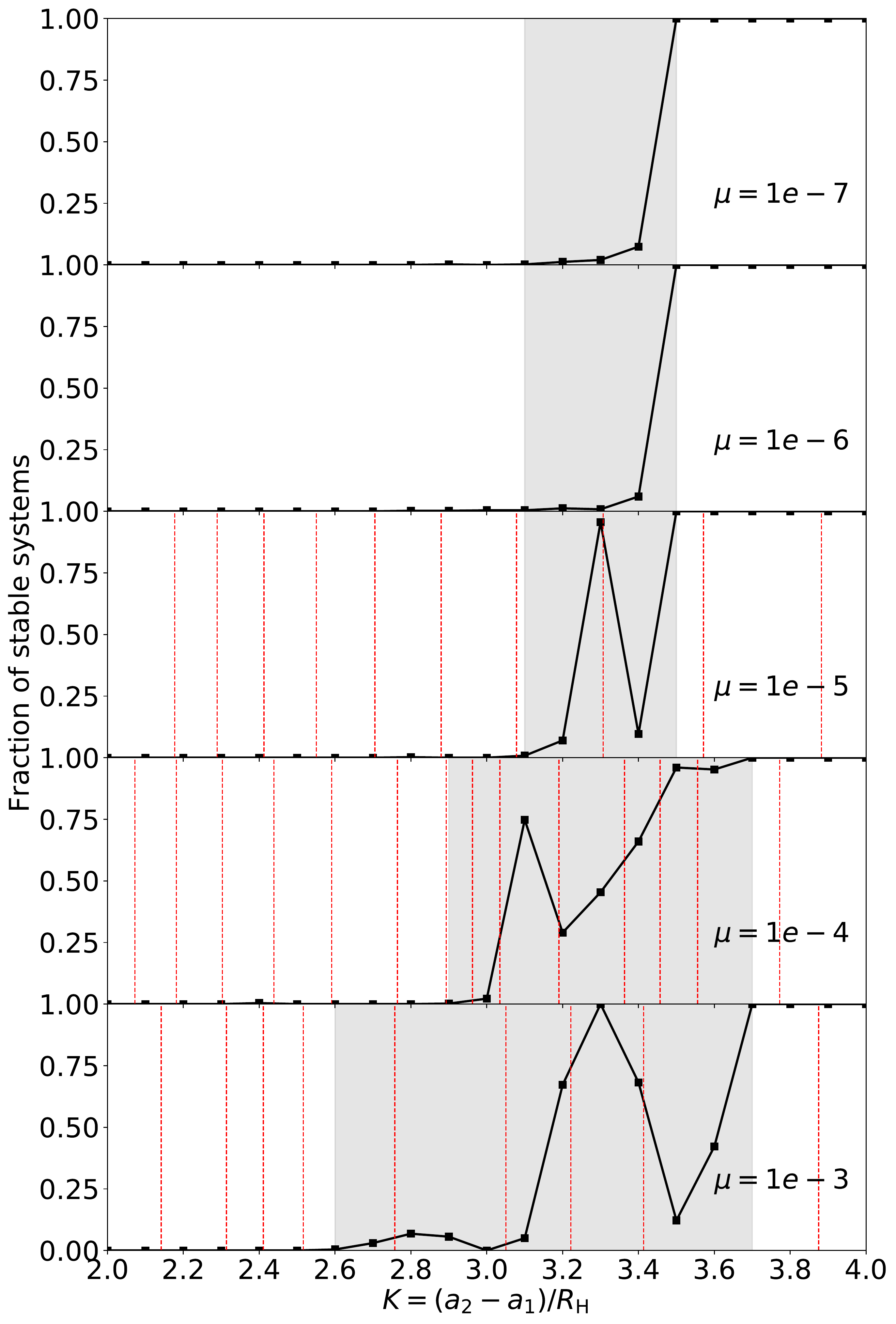}
    \caption{Fraction of stable runs for systems with different $\mu$ and initial $K$ in the ``planet + test mass'' simulations with no frictional force. The ``grey-zone'' regions start at the minimum $K$ with stable fraction $>1\%$ and end where the stable fraction reaches $100\%$. The dashed red lines mark the $K$ values that correspond to $(p+q):p$ mean-motion resonances for $p\leq20$ and $q\leq3$.
    }
    \label{fig:Hill-wRH-nf-fraction}
\end{figure}

Fig.~\ref{fig:Hill-wRH-nf-fraction} shows the fraction of stable runs in each suite of simulations. 
In all five panels (which present five different mass ratio $\mu$), there exist a ``grey zone'' of $K$ between unstable systems for small $K$'s and stable systems for large $K$'s.
The size of the ``grey zone'' depends on $\mu$.
Table~\ref{tab:Tinst-fit-RH} lists the values for the left and right boundary of the ``grey zone'', which we denote as $K_{\rm gz}$ and $K_{\rm crit}$, respectively.
Systems with $K < K_{\rm gz}$ have more than $99\%$ of chance to be unstable. 
Stability is guaranteed in the systems with $K \geq K_{\rm crit}$. 

In the ``grey zone'' for $\mu=10^{-5}$, $10^{-4}$ and $10^{-3}$, stable islands exist at $K \simeq 3.3$, $3.1$ and $3.3$, respectively. 
We find that these islands roughly correspond to the $K$ values of some mean-motion resonances (as marked in the figure).
No stable islands are found for $\mu=10^{-7}$ and $10^{-6}$ because their $R_{\rm H}$ is too small.

\begin{figure}
    \includegraphics[width=\columnwidth]{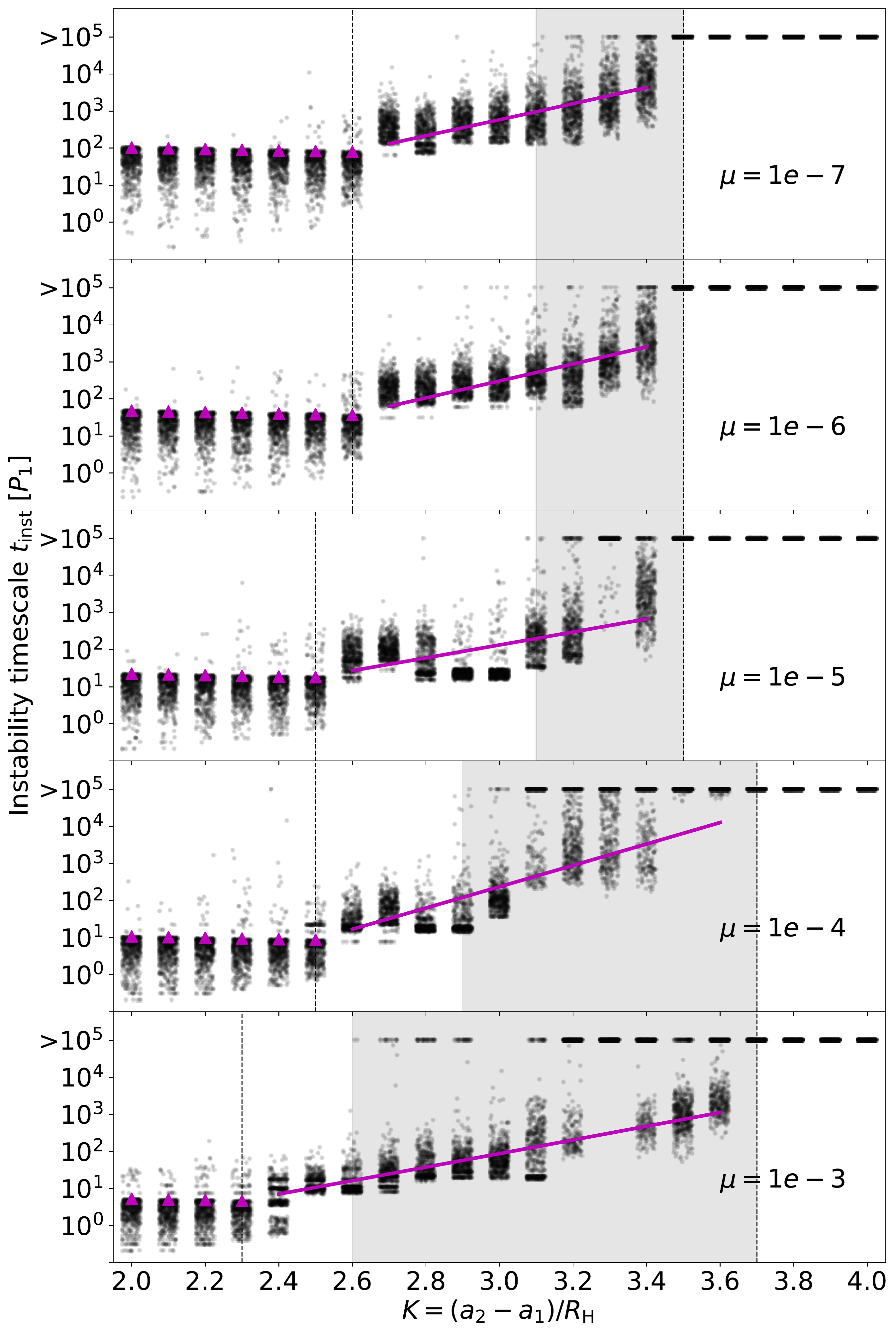}
    \caption{Instability timescale $t_{\rm inst}$ (black dots) for the simulations shown in Fig.~\ref{fig:Hill-wRH-nf-fraction}. The data points for each $K$ are manually spread on the horizontal axis for better display. The vertical dashed lines indicate $K_{\rm syn}$ and $K_{\rm crit}$, and the ``grey-zone'' regions are shaded. At $K\leq K_{\rm syn}$, the magenta triangles mark the synodic period between the test mass and the planet. The magenta lines show the fitting results using equation~\eqref{eq:Hill-wRH-Tinst-fit} in the transition regions between $K_{\rm syn}$ and $K_{\rm crit}$.
    }
    \label{fig:Hill-wRH-nf-Tinst}
\end{figure}

The ``degree of instability'' of a system can be characterised by $t_{\rm inst}$, the time for the instability to develop from two initially nearly circular orbits. 
Fig.~\ref{fig:Hill-wRH-nf-Tinst} shows the $t_{\rm inst}$ from our simulations. 
For each initial $K$, the distribution of $t_{\rm inst}$ results from the random selections of the initial argument of pericenter and mean anomaly. 
The horizontal spread is manually added for better display. 
The systems that remain stable at $10^5P_1$ are included in the plot and grouped in the $t_{\rm inst}>10^5P_1$ bins. 

When $K$ is small, $t_{\rm inst}$ is almost always less than the synodic period of the test mass from the planet,
\begin{equation}
\label{eq:Tsyn}
T_{\rm syn}(K) = \left(\frac{1}{P_1} - \frac{1}{P_2}\right)^{-1},
\end{equation} 
where $P_1$ and $P_2$ are the initial orbital periods of $m_1$ and $m_2$. 
This means that, except for a very small number of outliers, the mutual gravity between the planet and the test mass at their first orbital conjunction is strong enough to induce instability in a single shot. 
At least $80\%$ of the instabilities are single-shot when $K\leq K_{\rm syn} \sim 2.5$ (See Table~\ref{tab:Tinst-fit-RH}).  

Between $K_{\rm syn}$ and $K_{\rm crit}$ is a transition region, where the instability takes $t_{\rm inst} \gg T_{\rm syn}$ to develop. 
Note that $t_{\rm inst}$ can vary by one to two orders of magnitude for the same $\mu$ and $K$. 
However, the ``typical'' $t_{\rm inst}$ value is an exponential function of $K$. 
We fit the $\ln{(t_{\rm inst})}$-vs-$K$ relation with 
\begin{equation}
\label{eq:Hill-wRH-Tinst-fit}
\ln{(t_{\rm inst})} = \ln{(T_{\rm syn})} +  b\frac{K-K_{\rm syn}}{K_{\rm syn}}.
\end{equation}
using the least squares method. 
The data points with $t_{\rm inst}>10^{5}P_1$ are excluded from the fitting. 
Table~\ref{tab:Tinst-fit-RH} gives the fitting parameters for different values of $\mu$. 

\begin{table}
	\centering
    \caption{Parameters for the instability timescale in the ''planet + test mass'' systems (of different mass ratio $\mu=m_1/M$) with no frictional force. (i) $K_{\rm syn}$, the largest $K$ value that gives $t_{\rm inst}<T_{\rm syn}$ for 80\% of the runs; (ii) $K_{\rm gz}$, the minimum $K$ value to have $1\%$ systems to be stable (the left edge of the ``grey zone'' in Fig.~\ref{fig:Hill-wRH-nf-fraction}); (iii) $K_{\rm crit}$, the minimum $K$ value that ensures stability (the right edge of the ``grey zone'' in Fig.~\ref{fig:Hill-wRH-nf-fraction}); (iv) $b$, the best fit parameter $b$ in equation~\eqref{eq:Hill-wRH-Tinst-fit} with one standard deviation error. 
    }
    \label{tab:Tinst-fit-RH}
    \begin{tabular}{ ccccc } 
    \hline
     $\mu$ & $K_{\rm syn}$ &$K_{\rm gz}$ & $K_{\rm crit}$ & $b$\\
    \hline
     $10^{-7}$ & 2.6 & 3.1 & 3.5 & $13.8\pm0.1$\\
     $10^{-6}$ & 2.6 & 3.1 & 3.5 & $14.6\pm0.1$\\
     $10^{-5}$ & 2.5 & 3.1 & 3.5 & $10.9\pm0.1$\\
     $10^{-4}$ & 2.5 & 2.9 & 3.7 & $17.3\pm0.1$\\
     $10^{-3}$ & 2.3 & 2.6 & 3.7 & $10.39\pm0.05$\\ 
    \hline
    \end{tabular}
\end{table}

\subsection{Systems with frictional force}
\label{sec:Hill-wRH-wf}

Planets embedded in gaseous discs are subject to ``frictional'' forces that induce eccentricity damping and orbital migrations.
For example, a low-mass (non-gap openning) planet (of mass $m$ and semi-major axis $a$) experiences eccentricity damping on the timescale \citep[e.g.,][]{Tanaka2004}
\begin{align}
\nonumber
\tau & \sim   \frac{M^2h^4}{2\pi m \Sigma a^2}P \\
\nonumber
& \simeq
1.7\times10^3 \left(\frac{a}{\text{1AU}}\right)^{-2}
\left(\frac{\Sigma}{1700\text{g/cm}^2}\right)^{-1} \\
&\quad \times \left(\frac{h}{0.05}\right)^4\left(\frac{m}{m_{\oplus}}\right)^{-1}\left(\frac{M}{M_{\odot}}\right)^{2}P \label{eq:tau}
\end{align}
where $\Sigma$ is the disc surface density and $h$ is the aspect ratio, and we have adopted some representative numbers for protoplanetary discs. 
 
To assess the effect of the disc eccentricity damping to the dynamical instability, we re-run the simulations of Section~\ref{sec:Hill-wRH-nf} with an extra force \citep[see][]{Papaloizou2000},
\begin{equation}
\label{eq:wf-drag}
\vec{f} = -\frac{(\vec{v}\cdot\hat{r})\hat{r}}{\tau},
\end{equation}
applied to the test mass, where $\vec{v}$ is the test mass velocity and $\hat{r}$ is the unit radial vector.
For $e\ll1$, this force leads to eccentricity damping $\dot{e} \simeq -e/2\tau$.
We have also experimented with a different frictional force model with 
\begin{equation}
\label{eq:wf-drag-K}
\vec{f}= -\frac{\vec{v} - \vec{v}_{\rm K}}{\tau},
\end{equation}
where $\vec{v}_{\rm K}=\sqrt{GM/|\vec{r}|}\hat{\phi}$ is the Keplerian velocity at the location of the test mass. 
This force also gives $\dot{e} \simeq -e/2\tau$. 
We found that the effect of equation~\eqref{eq:wf-drag-K} on the stability of the ``planet+test particle'' system is similar to that of equation~\eqref{eq:wf-drag}.
In the following we describe our results based on equation~\eqref{eq:wf-drag}.

\begin{figure}
    \includegraphics[width=\columnwidth]{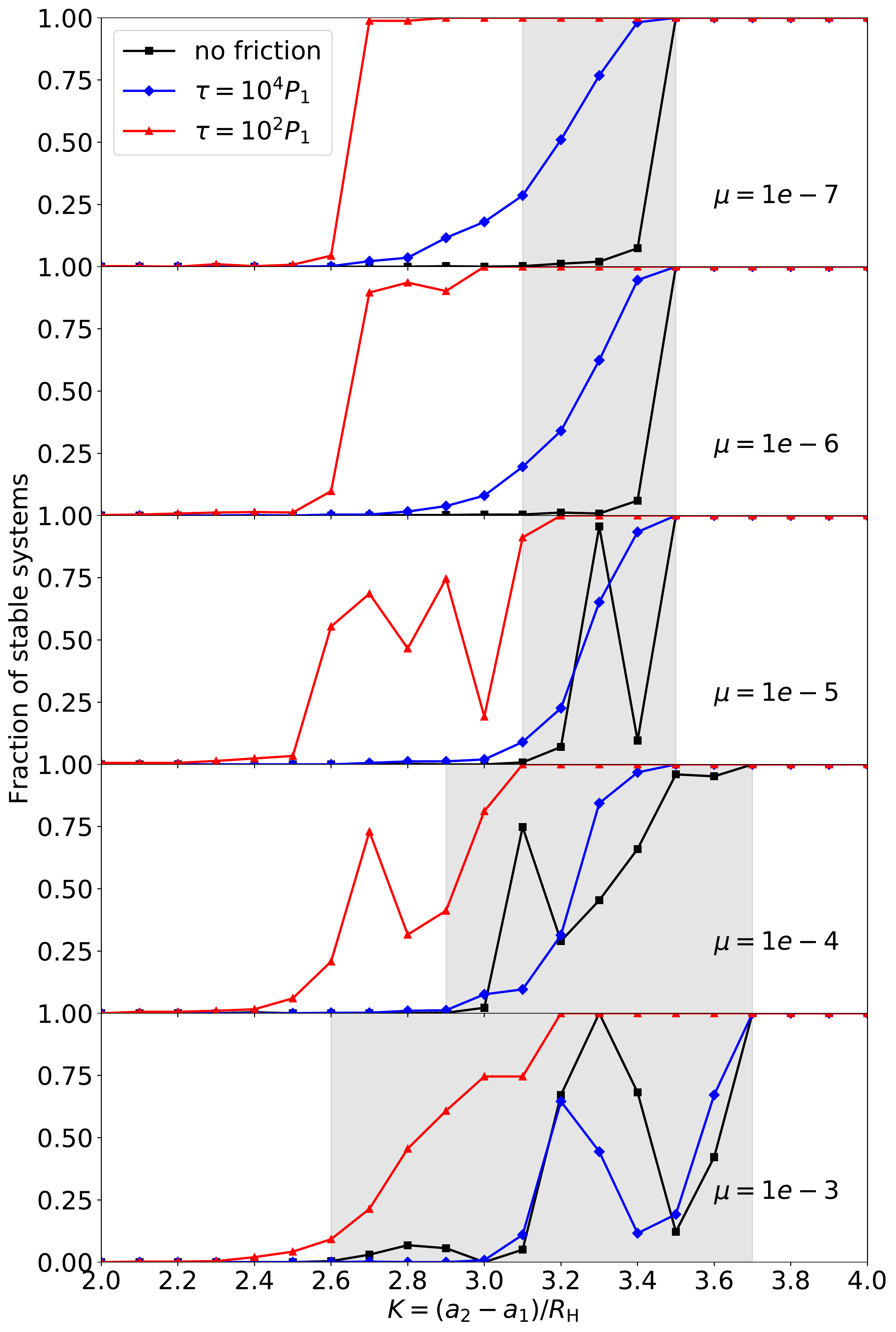}
    \caption{Same as Fig.~\ref{fig:Hill-wRH-nf-fraction}, except showing the results that includes the frictional force (equation~\ref{eq:wf-drag}) on the test particle.
    The red lines show the cases with $\tau=10^2P_1$ and the blue lines with $\tau=10^4P_1$. The black lines and the ``grey-zone'' regions are from the no-friction results (same as in Figure~\ref{fig:Hill-wRH-nf-fraction}).
    }
    \label{fig:Hill-wRH-wf-fraction}
\end{figure}

Fig.~\ref{fig:Hill-wRH-wf-fraction} shows the fraction of stable runs when the frictional force (equation~\ref{eq:wf-drag}) is applied.
The blue lines show the results from the weak friction case with $\tau=10^4P_1$.
The fraction of stable systems increases for all $K$ values except around the stable islands. 
The friction force can protect a system from instability by reducing the orbital eccentricity of the test mass.
However, the friction force can also affect the orbital energy of a test mass. 
Hence, inside the ``grey-zone'' stable islands, the protective effect of the mean motion resonances is largely weakened. 
Overall, as the stable fraction changes from 0 to 1 with increasing $K$, the weak friction force smooths the transition region by expanding the ``grey-zone'' to smaller values of $K$ and reducing the non-monotonic features inside the ``grey-zone''. 

The red curves in Fig.~\ref{fig:Hill-wRH-wf-fraction} show the results from the strong friction case with $\tau=10^2P_1$. 
The original ``grey zone'' is almost fully stabilised by the frictional force. 
The transition between ``always unstable'' and ``always stable'' now happens at smaller $K$'s.
New non-monotonic features appear, possibly because the transition zones now sweep through new resonances.

\begin{table}
	\centering
    \caption{Critical initial $K$ when the frictional force (equation~\ref{eq:wf-drag}) is applied in the ``planet + test mass'' systems for different the mass ratio $\mu=m_1/M$. (i) $K_{\tau=1e4}$, the largest $K$ value that allows instability when $\tau=10^4P_1$; (ii) $\widetilde{K}_{\tau=1e4}$, the critical $K$ when $\tau=10^4P_1$ as estimated by equation~\eqref{eq:Ktau-fit}; (iii) $K_{\tau=1e2}$, the largest $K$ that allows instability when $\tau=10^2P_1$; (iv) $\widetilde{K}_{\tau=1e2}$, the critical $K$ when $\tau=10^2P_1$ as estimated by equation~\eqref{eq:Ktau-fit}. 
    }
    \label{tab:K-fric-RH}
    \begin{tabular}{ccccc} 
    \hline
    $\mu$ & $K_{\tau=1e4}$ & $\widetilde{K}_{\tau=1e4}$ & $K_{\tau=1e2}$ & $\widetilde{K}_{\tau=1e2}$\\
    \hline
    $10^{-7}$ & 3.4 & 3.4 & 2.8 & 2.6 \\
    $10^{-6}$ & 3.4 & 3.4 & 2.9 & 2.8 \\
    $10^{-5}$ & 3.4 & 3.4 & 3.1 & 2.9 \\
    $10^{-4}$ & 3.4 & 3.6 & 3.0 & 2.9 \\ 
    $10^{-3}$ & 3.6 & 3.6 & 3.1 & 3.0 \\ 
    \hline
    \end{tabular}
\end{table}

From the fitting formula (equation~\ref{eq:Hill-wRH-Tinst-fit}), for each $\mu$ and $\tau$, there is a special initial orbital separation,
\begin{equation}
 K^* = \left[\frac{1}{b}\ln\left(\frac{\tau}{T_{\rm syn}}\right)+1\right] K_{\rm syn} \label{eq:Kstar}
\end{equation}
obtained by setting $t_{\rm inst}=\tau$.
We define
\begin{equation}
\widetilde{K}_{\tau} = 
\begin{cases}
     K^* & \text{for } K_\mathrm{syn} < K^* < K_\mathrm{crit}, \\
     K_\mathrm{syn} & \text{for } K^* < K_\mathrm{syn}, \\
     K_\mathrm{crit} & \text{for } K^* > K_\mathrm{crit}.
\end{cases}\label{eq:Ktau-fit}
\end{equation}
With initial $K>\widetilde{K}_{\tau}$, a system is expected to have $t_{\rm inst}>\tau$. 
That means the eccentricity damping is faster than the growth of instability and the system is expected to be stable.
Table~\ref{tab:K-fric-RH} gives the value of $\widetilde{K}_{\tau}$ from equation~\eqref{eq:Ktau-fit} and $K_{\tau}$, the true numerical value of the critical separation for $100\%$ stability.
Although equations~\eqref{eq:Hill-wRH-Tinst-fit} and~\eqref{eq:Ktau-fit} allow us to relate $K$ and the ``typical'' $t_{\rm inst}$, we note that even systems with the same initial $K$ can have a large spread of $t_{\rm inst}$'s (see Fig.~\ref{fig:Hill-wRH-nf-Tinst}).
Whether a system can be stabilised by friction depends more fundamentally on its $t_{\rm inst}$ than its initial $K$.   

\begin{figure}
    \includegraphics[width=\columnwidth]{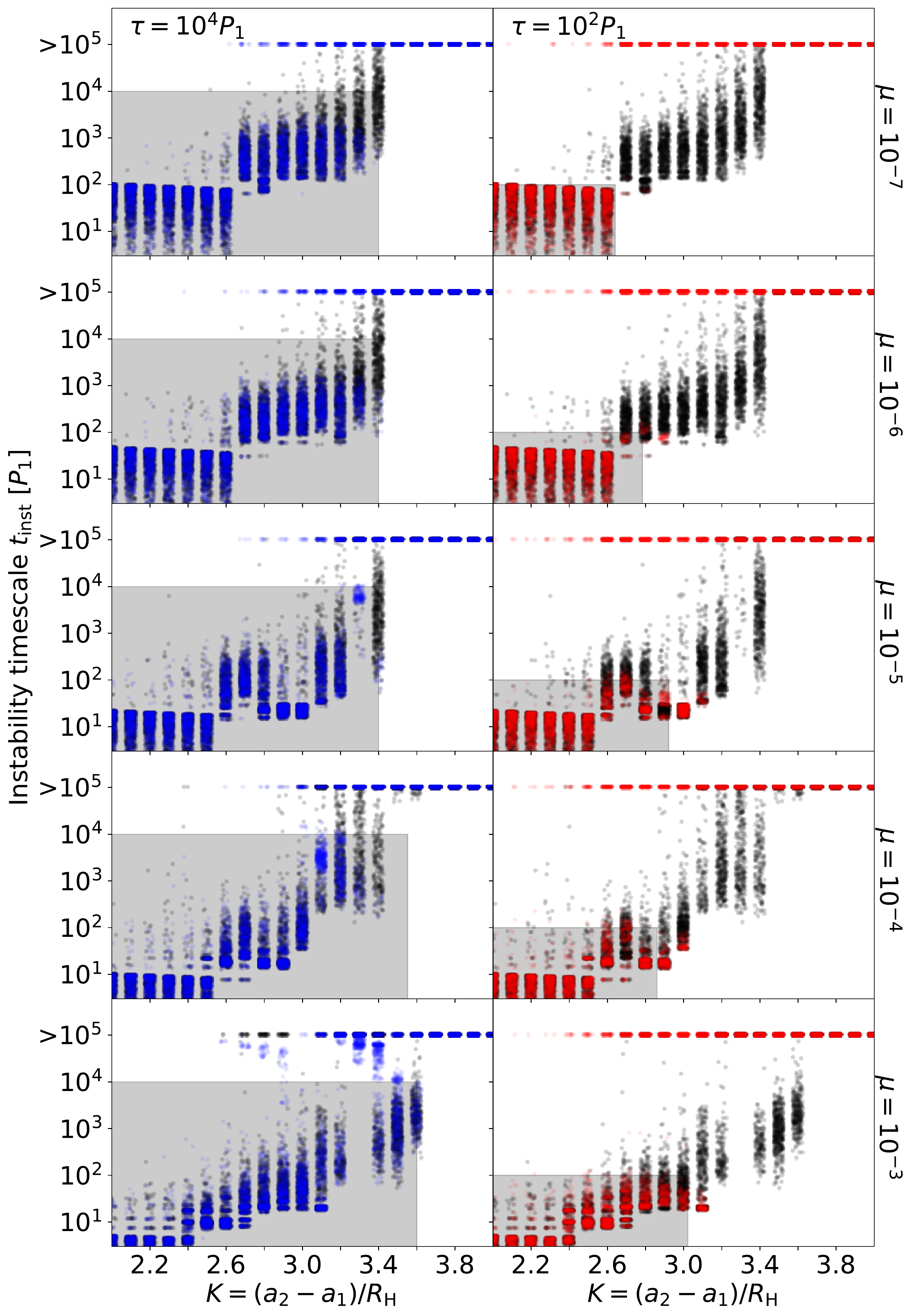}
    \caption{Instability timescale $t_{\rm inst}$ for ``planet + test mass'' systems with frictional force applied. The red dots show the results for the $\tau=10^2P_1$ case and the blue dots show the results for the $\tau=10^4P_1$ case. The black dots are the no-friction results (same as in Figure~\ref{fig:Hill-wRH-nf-Tinst}). The area within $[2.0,\widetilde{K}_{\tau}]\times[0,\tau]$ are grey shaded.
    }
    \label{fig:Hill-wRH-wf-Tinst}
\end{figure}

Fig.~\ref{fig:Hill-wRH-wf-Tinst} compares the $t_{\rm inst}$ results from the with-friction runs (coloured dots) to those from the no-friction runs (black dots). 
We shade the area with $K \in [2.0,\widetilde{K}_{\tau}]$ and $t_{\rm inst}\in[0,\tau]$.
Inside the shaded area, the instability growth is faster than the eccentricity damping ($t_{\rm inst}<\tau$), so the with-friction and no-frictions runs show the same distribution of $t_{\rm inst}$, and the coloured dots cover the black dots almost exactly.
Outside of the shaded area, the instability is weaker ($t_{\rm inst}>\tau$), so the systems can be stabilised by the frictional force; 
thus, the black dots are not covered by the coloured dots as the coloured dots are pushed upward to the simulation time limit at $t_{\rm inst}>10^{5}P_1$.

Due to the spread of $t_{\rm inst}$'s for runs with the same $K$, a damping force with a constant $\tau$ can only stabilise a fraction of the runs if $t_{\rm inst}(K)\sim\tau$.
That creates a wider region of ``grey zone'' where the outcome of the evolution is undetermined and a more smooth transition from ``all unstable'' to ``all stable'' (see Fig.~\ref{fig:Hill-wRH-wf-fraction}). 
The coloured outliers with $t_{\rm inst} \in (\tau,10^{5}P_1)$ in Fig.~\ref{fig:Hill-wRH-wf-Tinst} are systems that would be protected by mean motion resonances but is now destabilised by the damping force.

\section{A Linear Map Analysis of the Stability of the Restricted Three-body Problem}
\label{sec:map}

\subsection{Formalism of the linear map}
\label{sec:map-formalism}
In this section, we present an analytical map that can qualitatively reproduce the simulation results of Section~\ref{sec:Hill-wRH}. Our approach is modified and generalised from the previous work by \citet[hereafter DQT89]{Duncan1989}.

\begin{figure}
    \includegraphics[width=\columnwidth]{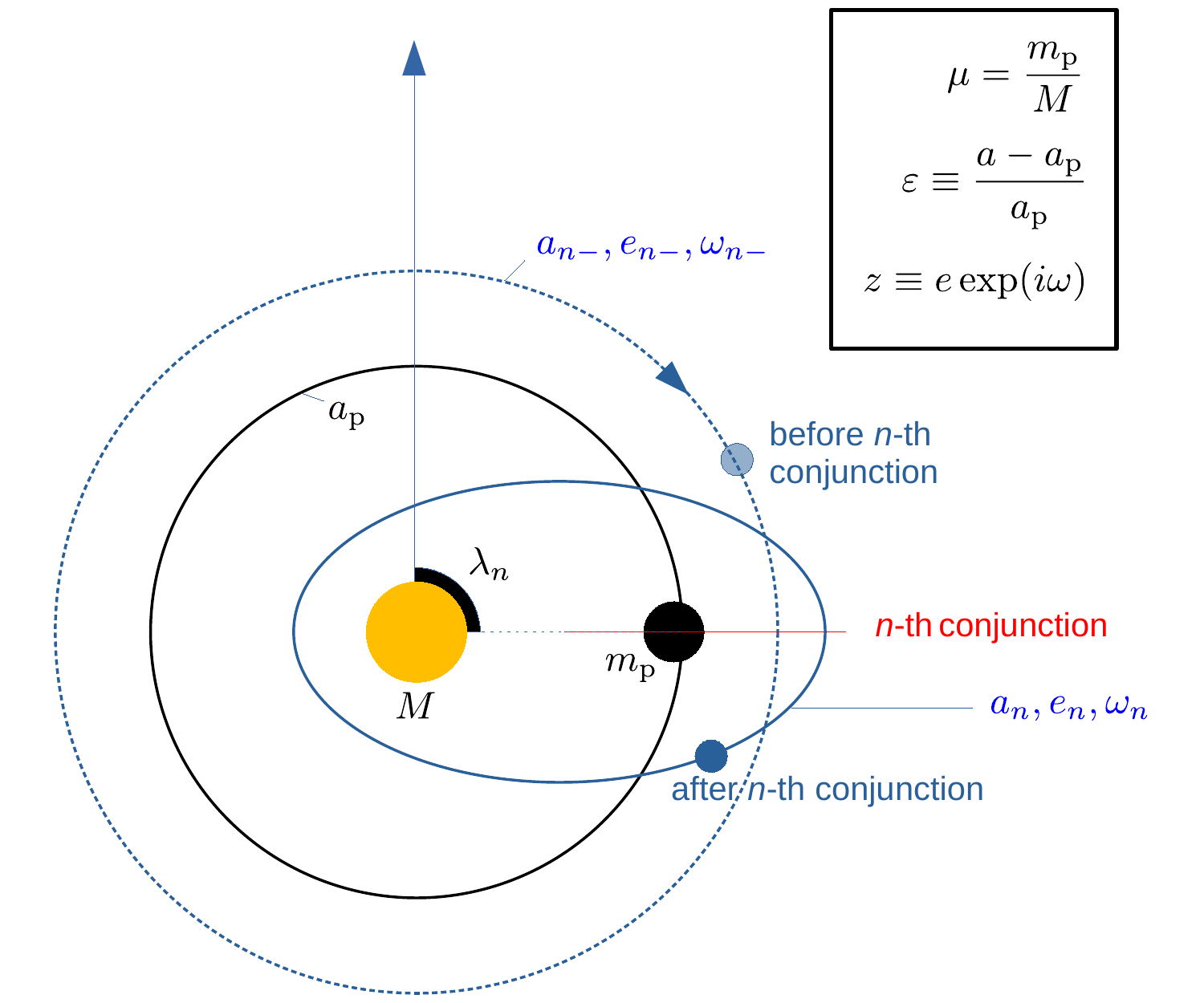}
    \caption{Setup of the linear map and the definition of the key parameters and variables.
    }
    \label{fig:map-notation}
\end{figure}

We map the evolution of the outer test particle's orbit to a succession of conjunctions with the inner planet. 
The leading-order change of the orbital elements of the test mass during a conjunction can be derived using the shearing box framework \citep{Henon1986}. 
We define
\begin{align}
	z &= e \exp\left(i\omega\right),\\
	\varepsilon &= \frac{a-a_p}{a_p},\\
	g &= \frac{8}{9} \left[2K_0\left(\frac{2}{3}\right)+K_1\left(\frac{2}{3}\right)\right] \approx 2.24,
\end{align}
where $K_0$ and $K_1$ are the modified Bessel functions of the second kind, and the subscript $p$ denotes the massive planet. 
We define $z_n$ to be the complex eccentricity just after the $n^{\rm th}$ conjunction, and $z_{n-}$ the complex eccentricity just before the $n^{\rm th}$ conjunction (see Fig.~\ref{fig:map-notation}). 
We use the same notations for the relative semi-major axis $\varepsilon$ and the longitude $\lambda$ of the test mass at the conjunction. 
During the $n^{\rm th}$ conjunction, the complex eccentricity is changed by the amount (see DQT89)
\begin{equation}
	\Delta z = \frac{i g \mu}{\varepsilon^2} e^{i\lambda_{n-}}, \label{eq:dz}
\end{equation}
and the following quantity (Jacobi integral) is conserved 
\begin{equation}
	3 \varepsilon - 4|z|^2. \label{eq:Jacobi}
\end{equation}
In DQT89, $\varepsilon$ is fixed to $\varepsilon_{n-}$ in equation~\eqref{eq:dz}, an approximation that becomes incorrect close to the instability boundary, where $\varepsilon$ can vary significantly. 
In this work, we use a leap-frog-like approach to evaluate $\varepsilon_{n,\mathrm{mid}}$, the average of the pre-conjunction and the estimated post-conjunction $\varepsilon$-values:
\begin{align}
	\varepsilon_{n,\mathrm{mid}} = \frac{1}{2}\left[\varepsilon_{n-} +  \sqrt{\varepsilon_{n-}^2+\frac{4}{3}\left(\left|z_{n-} + \frac{i g \mu}{\varepsilon_{n-}^2} e^{i\lambda_{n-}}\right|^2-|z_{n-}|^2\right)}\right].
\end{align}
The actual post-conjunction orbital elements are
\begin{align}
	z_{n} &= z_{n-} + \frac{i g \mu}{\varepsilon_{n,\mathrm{mid}}^2} e^{i\lambda_{n-}}, \label{eq:map-z-step}\\
	\varepsilon_{n} &= \sqrt{\varepsilon_{n-}^2+\frac{4\left(|z_{n}|^2-|z_{n-}|^2\right)}{3}}, \label{eq:map-eps-step}\\
	\lambda_{n} &= \lambda_{n-} + 2\pi |(1+\varepsilon_{n})^{\frac{3}{2}}-1|^{-1} \label{eq:map-lambda-step}.
\end{align}
In the no-damping case, the orbital elements are conserved between the conjunctions, i.e.
\begin{align}
	&z_{(n+1)-} = z_n, \label{eq:map-z-nf}\\
	&\varepsilon_{(n+1)-} = \varepsilon_n, \label{eq:map-eps-nf}\\
	&\lambda_{(n+1)-} = \lambda_n \label{eq:map-lambda-nf}.
\end{align}
In the damping case, with the frictional force described as in equation~\eqref{eq:tau} (corresponding to an exponential eccentricity damping with timescale $\tau$), we have at first order:
\begin{align}
	&z_{(n+1)-} = z_n \exp\left(-\frac{T_{\rm n}}{\tau}\right), \label{eq:map-z-wf}\\
	&\varepsilon_{(n+1)-} = \varepsilon_n, \label{eq:map-eps-wf}\\
	&\lambda_{(n+1)-} = \lambda_n \label{eq:map-lambda-wf},
\end{align}
where $T_n$ is the synodic period after the $n^{\rm th}$ conjunction. 
It can be derived from the orbital periods of the planet $P_p$ and test particle $P$:
\begin{equation}
	T_{\rm n} = \left(\frac{1}{P_{\rm p}}-\frac{1}{P}\right)^{-1} =\frac{P_{\rm p}}{1-(1+\varepsilon_n)^{-\frac{3}{2}}} \simeq \frac{2P_{\rm p}}{3\varepsilon_n} .
\end{equation}

Orbital crossings occur when $e > \varepsilon$ and equation~\eqref{eq:Hill-wRH-stop-condition} is met when $e\gtrsim\varepsilon-R_{\rm H}/a_{\rm p}$. However, both criteria lie outside the validity range of the map. 
In the following, we adopt the stability condition
\begin{equation}
	e_n < 0.5\varepsilon_n. \label{eq:stability}
\end{equation}
Empirical tests show that our results are insensitive to the precise numerical threshold ($0.5$ can be replaced by $0.4$ or $0.6$). 

Starting from $e=0$, equating $|\Delta z|$ (equation~\ref{eq:dz}) to $\varepsilon$ gives the Hill radius scaling ($\varepsilon \propto \mu^{1/3}$). 
Alternatively, equating the variable part of $|\lambda_n-\lambda_{n-}|$ to $\pi$ gives the resonance overlap criterion scaling \citep[$\varepsilon \propto \mu^{2/7}$,][DQT89, see Appendix~\ref{sec:map-appendix}]{Wisdom1980}. 
As we will see below however, these scalings do not account for the complexity of the instability threshold.

\subsection{No-friction map}

\subsubsection{Results}
\label{sec:map-results}

Using the algebraic map described in Section~\ref{sec:map-formalism}, we compute the ``eccentricity destabilisation time'' $t_{e}$, which is the time for a system to reach the limiting eccentricity to break equation~\eqref{eq:stability}.
A system is considered stable if it satisfies equation~\eqref{eq:stability} for 1000 conjunctions (equivalent to $10^3$--$10^5$ $P_1$ depending on the mass ratio and the initial $K$-value). 
The initial conjunction longitude is randomised, and the initial eccentricity of the test mass is set to $\mu$.

\begin{figure}
    \includegraphics[width=\columnwidth]{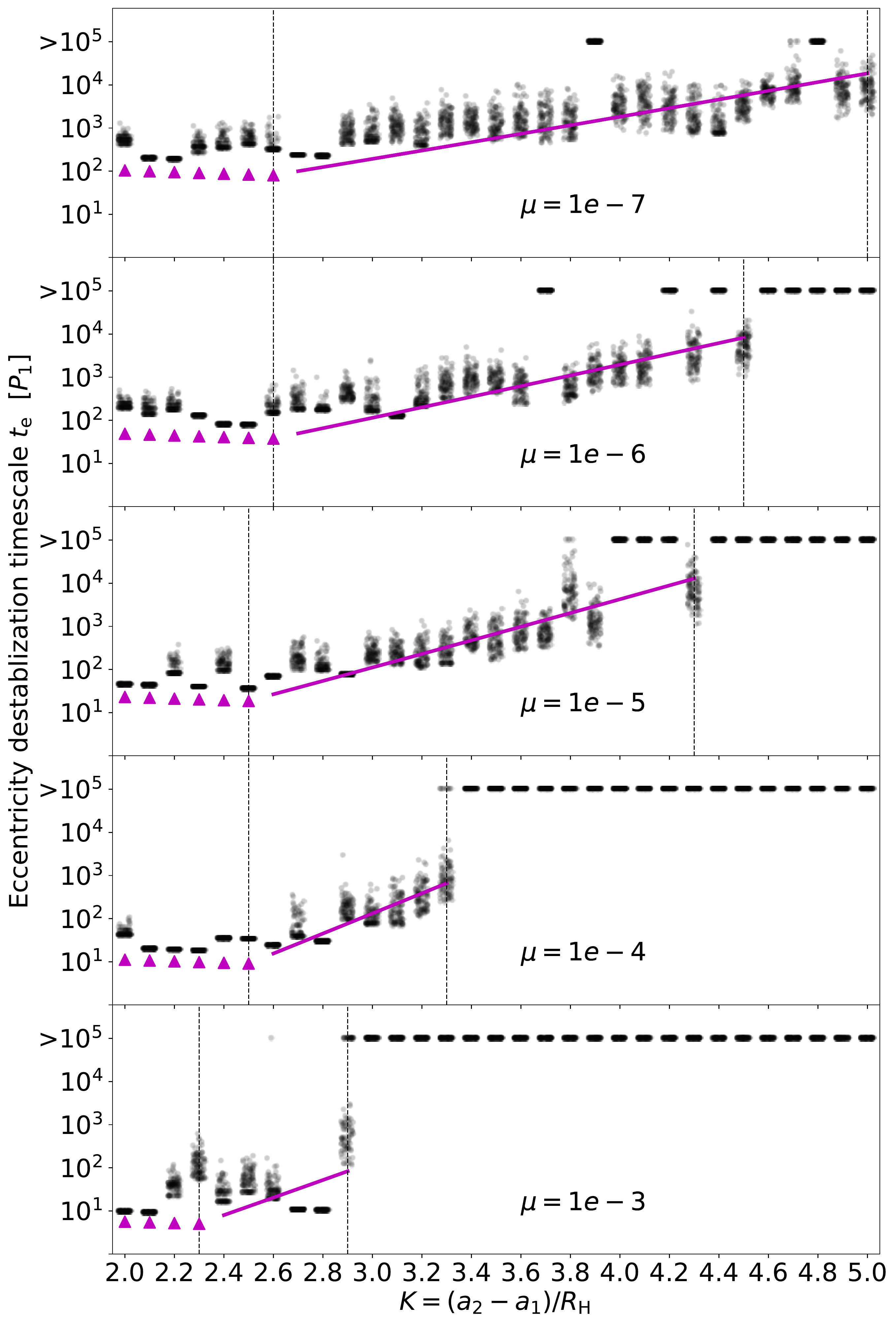}
    \caption{Similar to Fig.~\ref{fig:Hill-wRH-nf-Tinst}, except showing the eccentricity destabilisation timescale $t_e$ (black dots) from the map (see Section~\ref{sec:map-results}). No damping force is applied. The data points are manually spread on the horizontal axis for better display. The vertical dashed lines indicate $K\leq K_{\rm syn}$ and $K_{{\rm crit},e}$. The magenta triangles mark the synodic period between the test mass and the planet for $K_{\rm syn}$. The magenta lines show the fitting results of $t_e$ between $K_{\rm syn}$ and $K_{{\rm crit},e}$ using equation~\eqref{eq:Hill-map-Te-fit}. 
    }
    \label{fig:time-map}
\end{figure}

Fig.~\ref{fig:time-map} shows the results from the map without friction (i.e., with equations~\ref{eq:map-z-nf}-\ref{eq:map-lambda-nf}).
The behaviour of $t_e$ with respect to $K$ (for a given $\mu$) resembles that of $t_{\rm inst}$ from the $N$-body simulations. 
Fig.~\ref{fig:time-map} shows that the $t_e$-$K$ ``curve'' exhibits three branches.
The leftmost branch resembles the one-shot instability region in Fig.~\ref{fig:Hill-wRH-nf-Tinst}.
In this region, $t_e$ is very short and slightly decreases as $K$ increases. 
The rightmost region (except for the $\mu=10^{-7}$ case) covers $K > K_{{\rm crit},e}$, where $K_{\rm crit,e}$ is the largest $K$ that allows instability. 
The middle branch is the transition region, where the ``averaged'' $t_e$ increases exponentially with $K$.
Note that the actual $t_e$ has a wide spread for given $\mu$ and $K$; e.g., $t_e$ can differ by one to two orders of magnitude for two systems with the same $\mu$ and $K$.

Quantitatively, even if they are closely related, the $K_{{\rm syn},e}$ and $K_{{\rm crit},e}$ do not always equal to $K_{\rm syn}$ and $K_{\rm crit}$ (see Section~\ref{sec:Hill-wRH}) since the map and the $N$-body simulation use different instability criteria.
For the map, $t_e$ is always greater than or equal to the synodic period $T_{\rm syn}$ (see equation~\ref{eq:Tsyn}). 
We fit $t_e$ in the transition branch with 
\begin{equation}
\label{eq:Hill-map-Te-fit}
\ln{(t_e)} = \ln{(T_{\rm syn})} +  b_e\frac{K-K_{{\rm syn},e}}{K_{{\rm syn},e}},
\end{equation}
which has the same form as equation~\eqref{eq:Hill-wRH-Tinst-fit} for $t_{\rm inst}$. Table~\ref{tab:te-fit} gives the values of $K_{{\rm syn},e}$, $K_{{\rm crit},e}$ and $b_e$ for different $\mu$'s.

\begin{table}
	\centering
    \caption{Parameters for the transitional branch of the eccentricity destabilisation time from the no-friction map calculations: (i) $K_{\rm syn,e}$, same as the $K_{\rm syn}$ in Table~\ref{tab:K-fric-RH};
    (ii) $K_{{\rm crit},e}$, the largest $K$-value that allows instability; (iii) $b_e$, the best fit for the slope in equation~\eqref{eq:Hill-map-Te-fit} with one standard deviation errors. 
    }
    \label{tab:te-fit}
    \begin{tabular}{ cccc } 
    \hline
     $\mu$ & $K_{{\rm syn},e}$ & $K_{{\rm crit},e}$ & $b_e$\\
     \hline
     $10^{-7}$ & 2.6 & 5.0 & $6.6\pm0.1$\\
     $10^{-6}$ & 2.6 & 4.5 & $8.1\pm0.1$\\
     $10^{-5}$ & 2.5 & 4.3 & $9.8\pm0.1$\\
     $10^{-4}$ & 2.5 & 3.3 & $14.1\pm0.2$\\ 
     $10^{-3}$ & 2.3 & 2.9 & $11.5\pm0.3$\\ 
    \hline
    \end{tabular}
\end{table}

\subsubsection{Stability islands}

Similar to the $N$-body simulations (Section~\ref{sec:Hill-wRH}), the map evaluation also reveals some stability islands, such as at $K=3.9$ for $\mu=10^{-7}$ and $K=3.7$ for $\mu=10^{-6}$ (see Fig.~\ref{fig:time-map}).

To see why the stability islands exist, we analyse the map by assuming $g\mu/\varepsilon^2\ll e$. Using equation~\eqref{eq:map-z-step} and setting $z_{n-} = e\exp{i\omega}$, $\lambda_{n-} = \lambda$ and $\varepsilon_{n,\rm {mid}} = \varepsilon \simeq \varepsilon_n$, we find that the eccentricity change $\Delta e = |z_n|-|z_{n-}|$ in a conjunction is given by
\begin{equation}
\label{eq:Delta_e}
\Delta e \simeq  - \frac{g\mu}{\varepsilon^2}\sin(\lambda-\omega) .
\end{equation}
That means that $e$ will not change if the conjunction happens exactly when the test mass is at its pericenter or apocenter ($\lambda-\omega = 0$ or $\pi$).
When the planet and the test particle are at conjunction, the mean longitude of the two objects are the same. 
Hence, $\phi\equiv\lambda-\omega$ corresponds to the resonance angle of the planet and the test mass if they are in a first-order two-body mean motion resonance.
The cumulative change of $e$ will be zero if $\phi$ ``librates'' (discretely at each conjunction) around $0$ or $\pi$.

From equations\eqref{eq:map-z-step} and \eqref{eq:map-lambda-step} we find that the changes in $\omega$ and $\lambda$ in a conjunction are given by
\begin{align}
\Delta \omega &\simeq
\frac{g\mu}{e\varepsilon^2}\cos\phi, \\
\Delta \lambda &\simeq \text{wrap}\left(\frac{4\pi}{3\varepsilon}\right) \label{eq:map-delta-lambda},
\end{align}
where we wrap large angles into $[0,2\pi)$. 
If $e$ is a small constant because $\phi$ remains close to $0$ or $\pi$, equation~\eqref{eq:map-eps-step} suggests that $\varepsilon$ is almost a constant. 
Hence, we get
\begin{equation}
\Delta \phi = \text{wrap}\left(\frac{4\pi}{3\varepsilon}\right) - \frac{g\mu}{e\varepsilon^2}\cos\phi,
\end{equation}
which means the map allows $\Delta e$ and $\Delta \phi$ to be zero at the same time for some specific value of $\varepsilon$. 
The equilibrium is stable ($\partial^2 \Delta \phi / \partial \phi^2 <0$) at $\phi=\pi$, allowing a long-term ``libration'' of $\phi$ around $\phi=\pi$. 
This is why the systems with some special initial $K$ can have long-term stability according to the map.

\begin{figure}
    \includegraphics[width=\columnwidth]{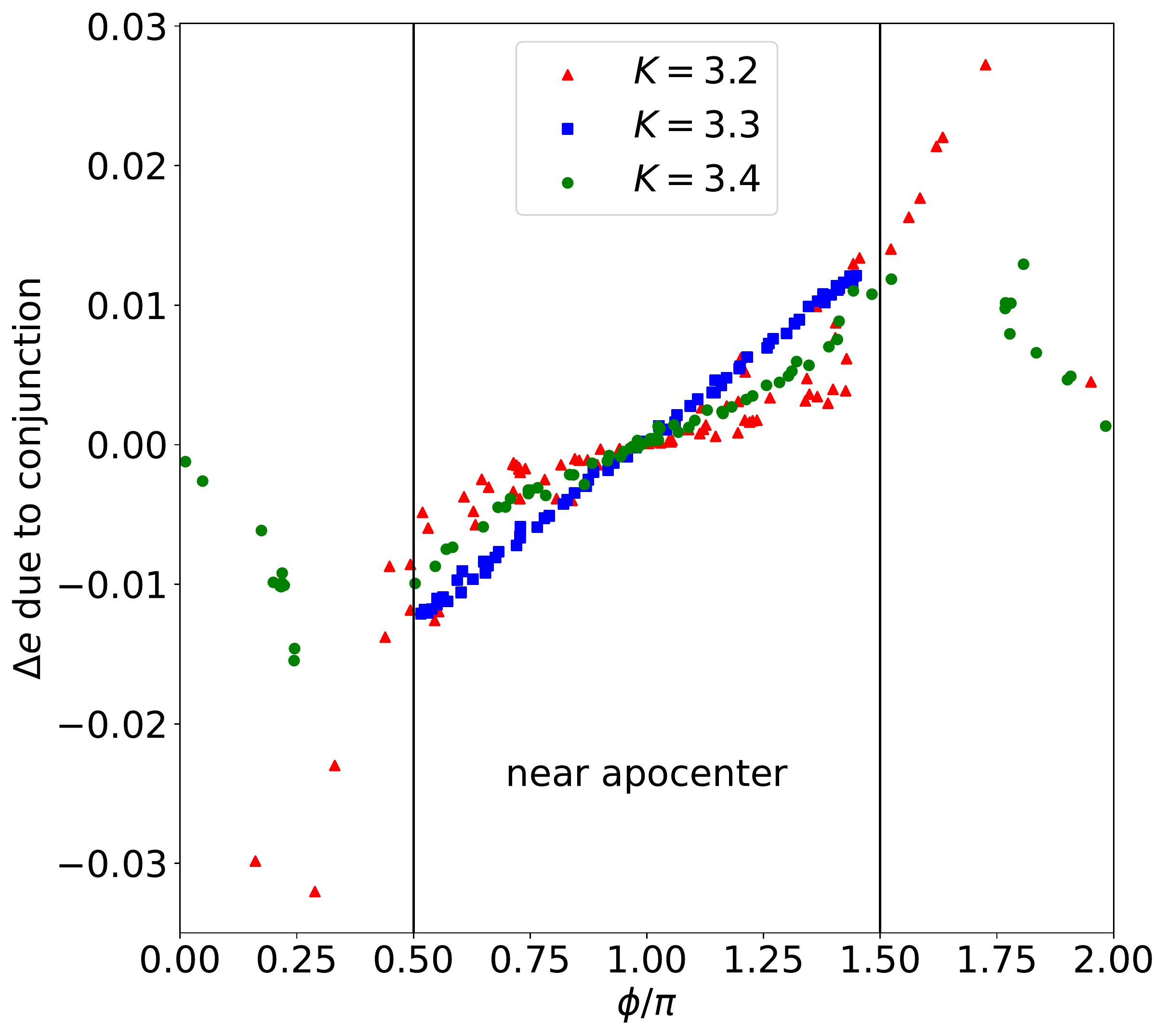}
	\caption{Distribution of $\phi=\lambda-\omega$ and $\Delta e$ at conjunctions in the $N$-body simulations with $\mu=10^{-5}$ and $K=3.2$ (red), $3.3$ (blue) and $3.4$ (green). This suite of $N$-body simulations finds a stability island at $K=3.3$. Only the conjunctions during the first 1000 orbits are shown.}
	\label{fig:map-island}
\end{figure}

This stabilising mechanism can also be applied to the $N$-body simulations and explain the stable islands found in Section~\ref{sec:Hill-wRH} (see Figs.~\ref{fig:Hill-wRH-nf-fraction} and \ref{fig:Hill-wRH-nf-Tinst}).
Fig.~\ref{fig:map-island} shows the distribution of $\phi$ and $\Delta e$ at conjunctions in the ``planet + test mass'' $N$-body simulations. 
We select the cases with $\mu=10^{-5}$ and $K=3.2$ (red), $3.3$ (blue) and $3.4$ (green), where $K=3.3$ is in a stable island while $K=3.2$ and $3.4$ are outside of the island.
All simulated conjunctions approximately follow $\Delta e \propto -\sin\phi$ as equation~\eqref{eq:Delta_e} predicts.
Near the island (blue), $\phi$ always stays between $\pi/2$ and $3\pi/2$ and distributes symmetrically around $\phi=\pi$, which agrees with the stabilising mechanism derived using the map.

\begin{figure}
    \includegraphics[width=\columnwidth]{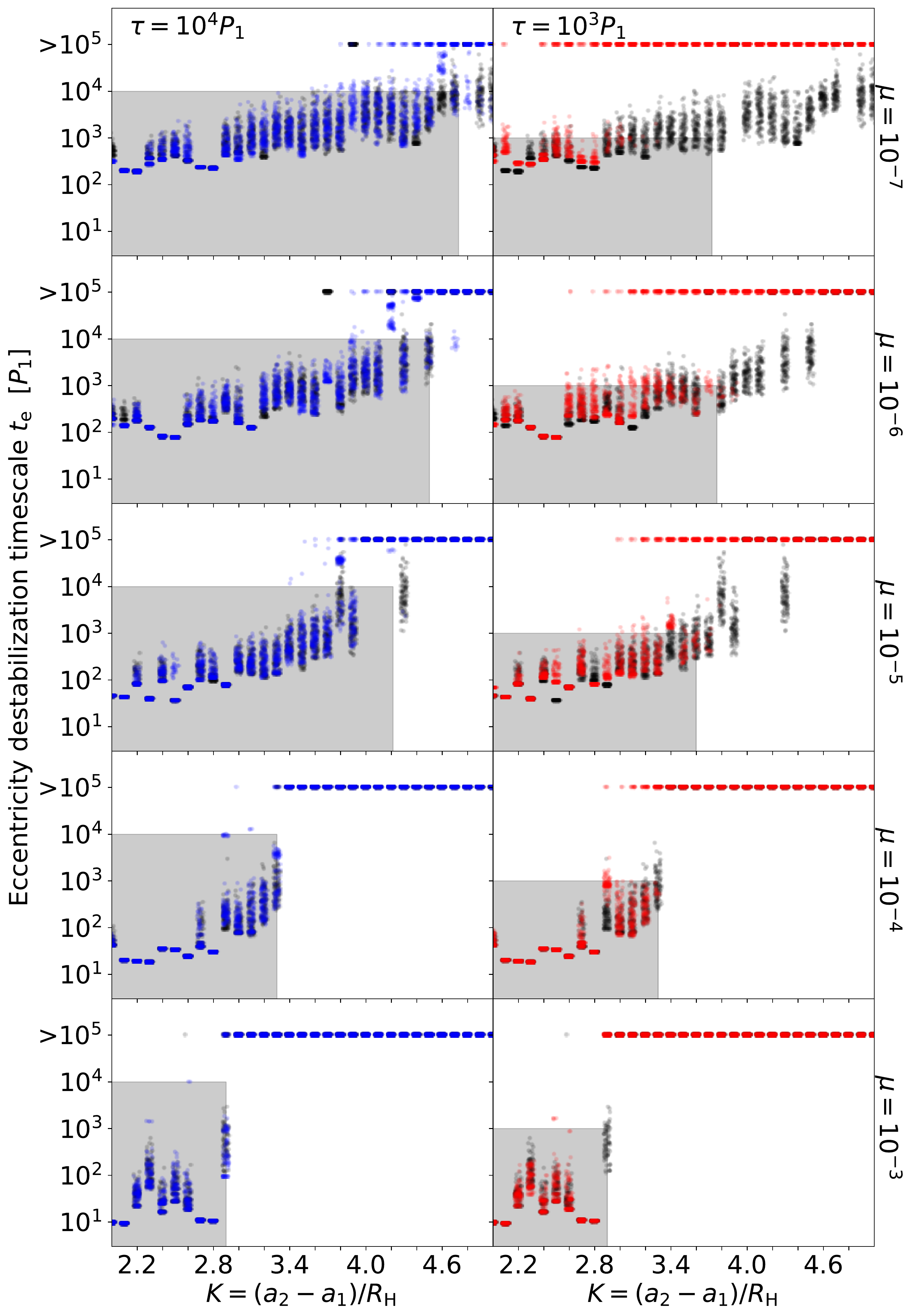}
    \caption{Similar to Fig.~\ref{fig:Hill-wRH-wf-Tinst}, but showing the eccentricity destabilisation timescale $t_e$ (which is the time when equation~\ref{eq:stability} breaks down) from the map. 
    }
    \label{fig:time-map-wf}
\end{figure}

\subsection{Map with friction}

Fig.~\ref{fig:time-map-wf} shows the eccentricity destabilisation timescale $t_e$ from the map with the frictional damping included (see equation~\ref{eq:map-z-wf}) for two damping timescale $\tau=10^4P_1$ and $\tau=10^3P_1$. 
We have also experimented with $\tau = 10^2P_1$ and found that all runs are stable.
Similar to equations~\eqref{eq:Kstar} and \eqref{eq:Ktau-fit} for the $N$-body simulations, we define the special initial spacing $K_e^*$ via $t_e(K) \simeq \tau$:
\begin{equation}
\label{eq:Ktaue-fit}
{K}_{e}^*=\left[\frac{1}{b}\ln\left(\frac{\tau}{T_{\rm syn}}\right)+1\right]K_{{\rm syn},e}.
\end{equation}
Following equation~\eqref{eq:Hill-map-Te-fit}, we use $K_e^*$ to derine $\widetilde{K}_{\tau,e}$, the map estimate of the largest $K$-value that allows instability when an eccentricity damping of timescale $\tau$ is applied.
In Fig.~\ref{fig:time-map-wf}, we shade the area with $K\in[2.0,\widetilde{K}_{\tau,e}]$ and $t_e \in [0,\tau]$, where the instability is expected to be stronger than the frictional damping. 
The result shows that the eccentricity can only grow in systems inside the shaded area. 
Outside of the shaded area, where $t_e>\tau$, the damping is too strong for the eccentricity to grow.

\begin{figure}
	\includegraphics[width=\columnwidth]{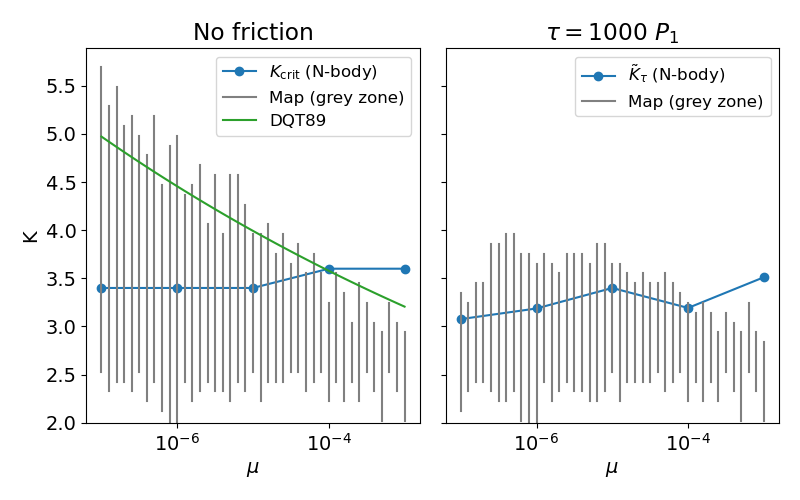}
	\caption{Transition zone between stability (all particles are stable) and instability (particles are unstable in less than two synodic periods) as a function of the mass ratio $\mu$. For each $\mu$, 100 test particles are evolved for 1000 conjunctions using the map. The green line is a fit to the end of the transition zone without friction ($K\propto \mu^{2/7}$, as obtained by DQT89). The blue line is the predicted effect of the friction using the $N$-body simulations (Table~\ref{tab:Tinst-fit-RH} and equation~\ref{eq:Ktau-fit}). The discrepancy is large between the $N$-body and map results for the no-friction case (left), mostly due to the different criteria for instability. On the other hand, the results are much more similar when we add friction, as both versions of instability behave alike.}
	\label{fig:K-map-vs-N}
\end{figure}

Fig.~\ref{fig:K-map-vs-N} compares the size of the transition zone $[K_{{\rm syn},e}, K_{{\rm crit},e}]$ from the map to $K_{\rm crit}$ from the $N$-body simulation (see Tables~\ref{tab:Tinst-fit-RH} and \ref{tab:K-fric-RH}) in the no-friction and friction case.
Better agreement between the stability thresholds is found when the damping is applied. 
Although the map does not exactly reproduce the results of the $N$-body simulations, it correctly predicts the orders of magnitude of the instability timescale and the effect of friction on stability. 
The grey zone in Fig.~\ref{fig:K-map-vs-N} is impacted by the damping term in the same way as the effect of frictions in the $N$-body simulations (see Section~\ref{sec:Hill-wRH-wf}).
The advantage of the map is twofold: it is much quicker to compute than $N$-body simulations, which allows a better sampling of the parameter space, and its constitutive equations shed some light on the instability mechanisms.

\section{Instability of Multi-Planet systems with Frictions}
\label{sec:NP}

In this section, we consider systems with multiple (non-zero mass) planets. 
Our simulated systems consist of a central star with mass $M$ and planets with mass $m_j= 2^{2-j}\mu M$ for $j=1,2,..,N$. The initial orbital separation between two neighbouring planets is set as 
\begin{equation}
a_{j+1} - a_{j} = KR_{\rm H},
\end{equation}
where $K$ the same dimensionless orbital separation for all neighbouring pairs and
\begin{equation}
R_{\rm H} \equiv \frac{a_{j}+a_{j+1}}{2} \left(\frac{m_{j}+m_{j+1}}{M}\right)^{1/3}
\end{equation}
is the mutual Hill radius of $m_{j}$ and $m_{j+1}$. The innermost (also the most massive) planet is given initial eccentricity $e_1=0$ and other planets ($j\geq2$) have initial eccentricities $e_j=10^{-5}$. 
All planets are co-planar and have zero mutual inclination.
For some experiments, we apply equation~\eqref{eq:wf-drag} to all planets to model the frictional damping. 

Same as in the previous sections, we consider different combinations of $\mu$, $K$ and $\tau$. 
We carry out 50 runs for each combination with randomised initial angles, anomalies and longitudes. 
Each simulation runs up to $10^{6}P_1$.

It has been well recognised that stability in two-planet systems and that in the systems with more than two planets are substantially different:
the Hill stability criterion only applies to two-planet systems; systems with three or more planets can become unstable even when the initial planetary separations are large, although the instability growth time increase rapidly with the separation \citep[see][and refs therein]{Pu2015}.
Hence, we will discuss the results from the simulations with $N=2$ and $N=3$ separately.

\subsection{Two-planet systems}

We set up the simulations with $\mu=10^{-6}$, $10^{-5}$, $10^{-4}$, $10^{-3}$ and initial $K=2.0$, $2.1$, $2.2$, $...$, $4.0$, and repeat all suite three times for ``no-friction'', $\tau=10^2P_1$, and $\tau=10^4P_1$.

\begin{figure}
    \includegraphics[width=\columnwidth]{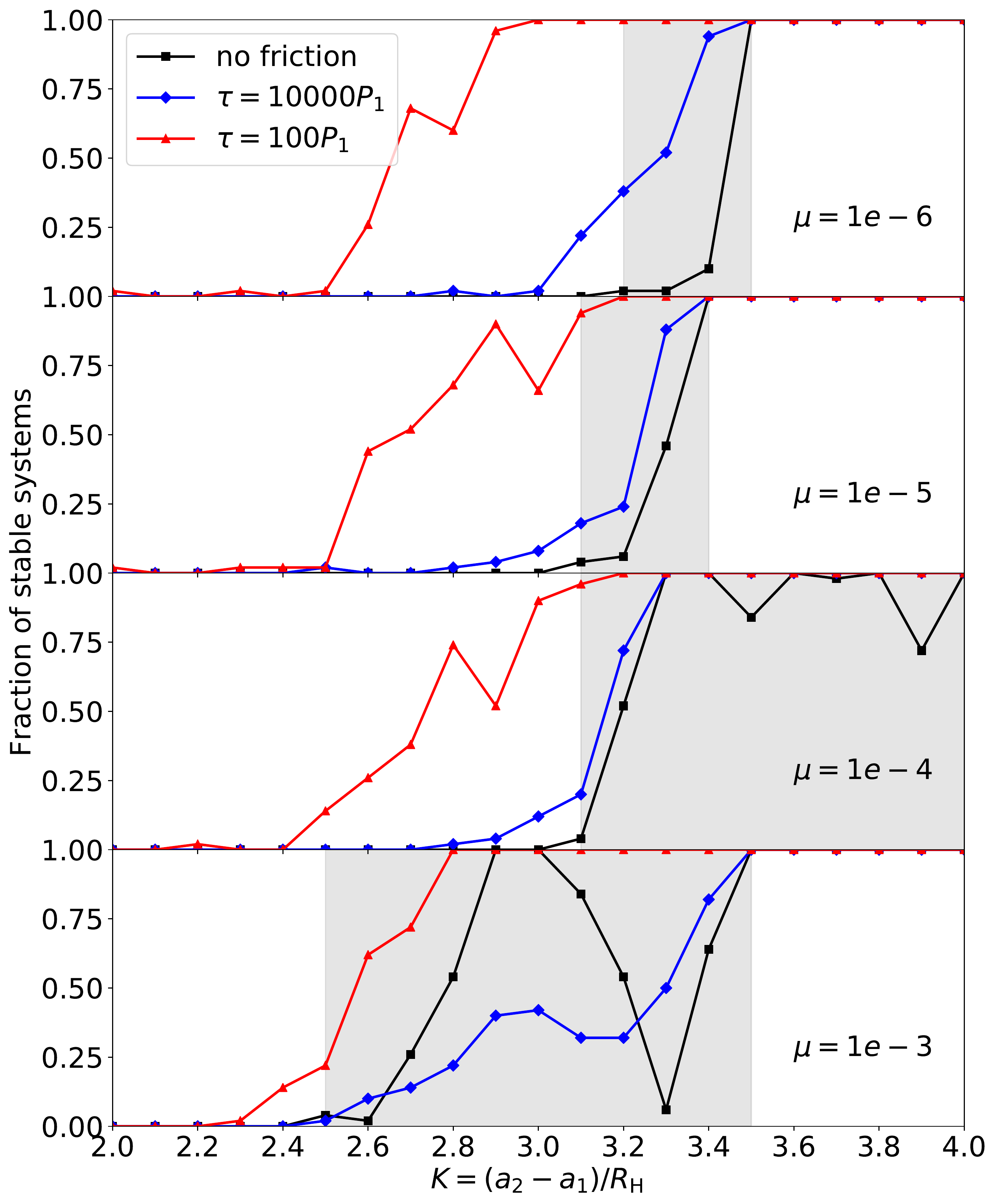}
    \caption{Similar to Fig.~\ref{fig:Hill-wRH-wf-fraction}, except showing the fraction of stable runs for systems with two finite-mass planets after $10^6P_1$. The black lines are from the ``no-friction'' runs. The red lines show the cases with $\tau=10^2P_1$ and the blue lines show the cases with $\tau=10^4P_1$. 
    }
    \label{fig:frac-N2}
\end{figure}

Fig.~\ref{fig:frac-N2} shows the fraction of stable runs in the two-planet simulations. 
We give the values of $K_{\rm gz}$ and $K_{\rm crit}$ for the ``grey zone'' (where a system can be either stable or unstable) in the ``no-friction'' runs in Table~\ref{tab:Tinst-fit-2P}.
Unlike the test-mass case (see Fig.~\ref{fig:Hill-wRH-nf-fraction}), the ``grey zones'' for two-planet systems do not contain any stability island, except at $K=2.7$ to $3.2$ for $\mu=10^{-3}$.
The strong instability at $K\sim3.3$ for $\mu=10^{-3}$ almost coincides with the location of the $5:3$ mean-motion resonance in the system.
When the friction is applied, the stable fraction increases almost everywhere, and decreases inside the stability islands for for $\mu=10^{-3}$.

\begin{figure}
    \includegraphics[width=\columnwidth]{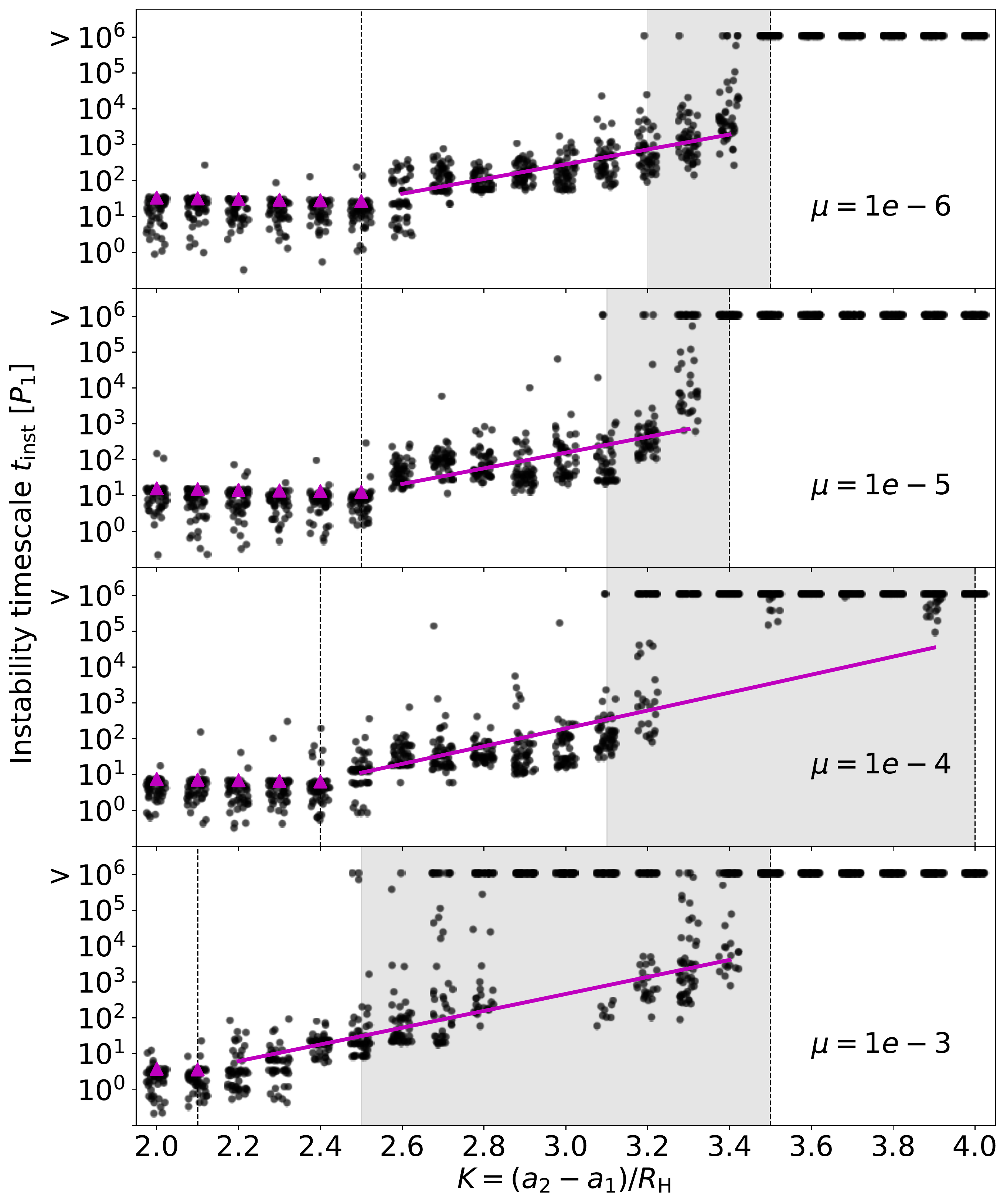}
    \caption{Similar to Fig.~\ref{fig:Hill-wRH-nf-Tinst}, except showing the instability timescale $t_{\rm inst}$ (magenta dots) for systems with two finite-mass planets. No friction force is applied. The vertical dashed lines indicate $K_{\rm syn}$ and $K_{\rm crit}$, and the ``grey-zone'' regions are shaded. At $K\leq K_{\rm syn}$, the magenta triangles mark the synodic period between the the two planets. The magenta lines show the fitting formulae using equation~\eqref{eq:Hill-wRH-Tinst-fit} in the transition region (which span from $K_{\rm syn}$ to $K_{\rm crit}$).
    }
    \label{fig:N2-nf-Tinst}
\end{figure}

Fig.~\ref{fig:N2-nf-Tinst} shows the instability growth time, $t_{\rm inst}$, from the no-friction simulations (with the systems that remain stable in $10^6P_1$ being added as the $t_{\rm inst}>10^6P_1$ dots). 
Similar to the test-mass cases (see Fig.~\ref{fig:Hill-wRH-nf-Tinst}), when $K$ is small ($\leq K_{\rm syn}$ given in Table~\ref{tab:Tinst-fit-2P}), $t_{\rm inst}$ is almost always less than the synodic period (equation~\ref{eq:Tsyn}) of the two planets:
the mutual gravity between the planets is strong enough to trigger immediate instability at their first orbital conjunction.  

Between $K_{\rm syn}$ and $K_{\rm crit}$, the instability growth time $t_{\rm inst}$ can vary by several orders of magnitude for the same $\mu$ and $K$. 
The range of the instability time is roughly an exponential function of $K$, despite the two outliers in the $\mu=10^{-4}$ case at $K=3.5$ and $3.9$.
We fit the data points in the transition region that has $t_{\rm inst}<10^{6}$ with equation~\eqref{eq:Hill-wRH-Tinst-fit}. 
Table~\ref{tab:Tinst-fit-2P} lists the fitting results.

\begin{figure}
    \includegraphics[width=\columnwidth]{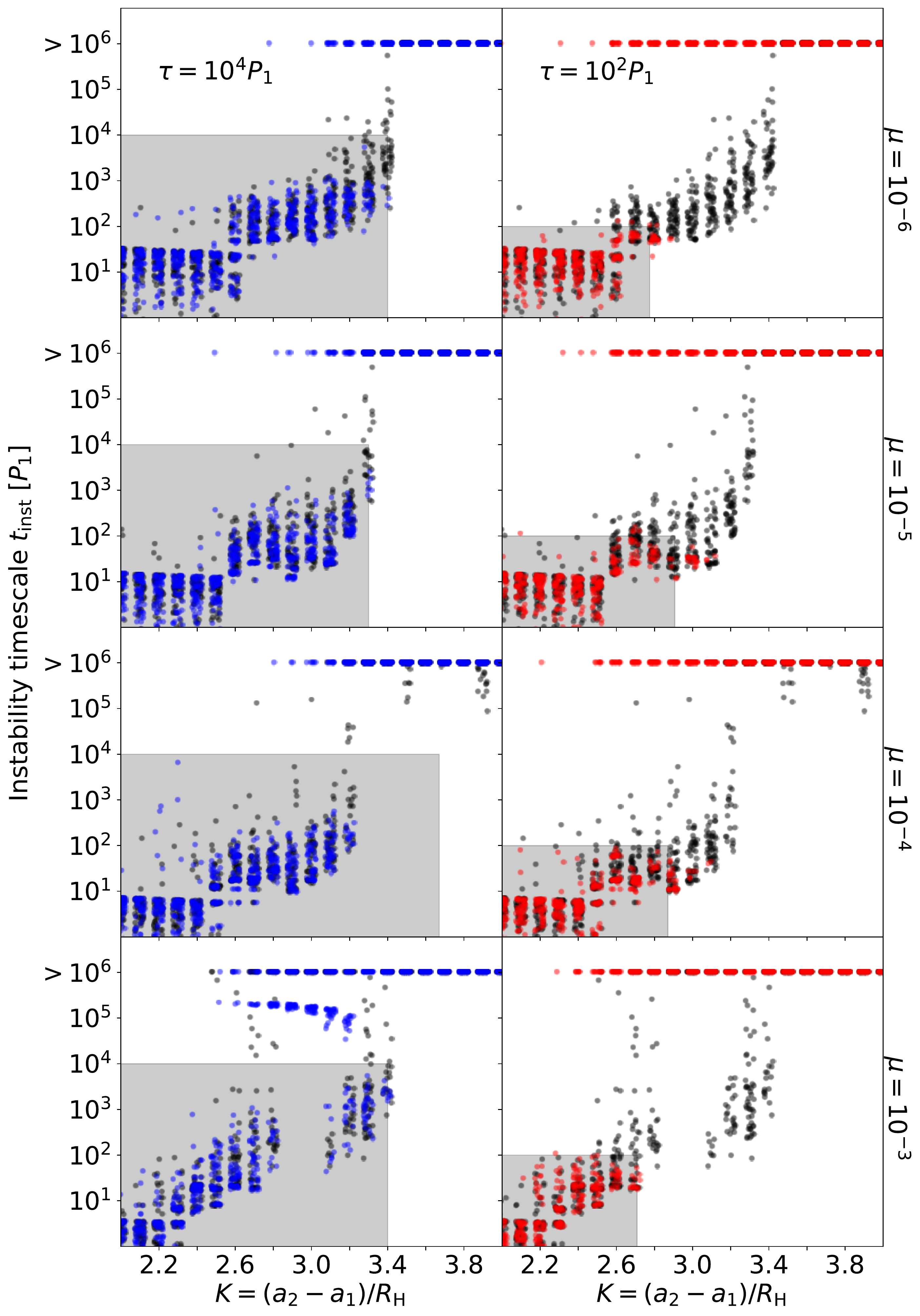}
    \caption{Similar to Fig.~\ref{fig:Hill-wRH-wf-Tinst}, but showing the instability time $t_{\rm inst}$ for two-planet systems with frictional force applied. The red dots show the results from the $\tau=10^2P_1$ cases and the blue dots show those from the $\tau=10^4P_1$ cases. The black dots are from the no-friction runs (same as in Figure~\ref{fig:N2-nf-Tinst}). The area within $[2.0,\widetilde{K}_{\tau}]\times[0,\tau]$ are grey shaded.
    }
    \label{fig:N2-wf-Tinst}
\end{figure}

\begin{table}
	\centering
    \caption{Parameters for the instability timescale in the two-planet systems (of different mass ratio $\mu=m_2/M$) with no frictional force. (i) $K_{\rm syn}$, the largest $K$ value that gives $t_{\rm inst}<T_{\rm syn}$ for 80\% of the runs; (ii) $K_{\rm gz}$, the minimum $K$ value to have $1\%$ systems to be stable (the left edge of the ``grey zone'' in Fig.~\ref{fig:Hill-wRH-nf-fraction}); (iii) $K_{\rm crit}$, the minimum $K$ value that ensures stability (the right edge of the ``grey zone'' in Fig.~\ref{fig:Hill-wRH-nf-fraction}); (iv) $b$, the best fit parameter $b$ in equation~\eqref{eq:Hill-wRH-Tinst-fit} with $1\sigma$ error. 
    }
    \label{tab:Tinst-fit-2P}
    \begin{tabular}{ ccccc } 
    \hline
     $\mu$ & $K_{\rm syn}$ & $K_{\rm gz}$ & $K_{\rm crit}$ & $b$\\
    \hline
     $10^{-6}$ & 2.5 & 3.2 & 3.5 & $12.7\pm0.2$\\
     $10^{-5}$ & 2.5 & 3.1 & 3.4 & $13.4\pm0.4$\\
     $10^{-4}$ & 2.4 & 3.1 & 4.0 & $14.3\pm0.4$\\ 
     $10^{-3}$ & 2.1 & 2.5 & 3.5 & $12.0\pm0.3$\\ 
    \hline
    \end{tabular}
\end{table}

\begin{table}
	\centering
    \caption{Critical initial $K$ when the frictional force (equation~\ref{eq:wf-drag}) is applied in the two-planet systems for different the mass ratio $\mu=m_2/M$. (i) $K_{\tau=1e4}$, the largest $K$ value that allows instability when $\tau=10^4P_1$; (ii) $\widetilde{K}_{\tau=1e4}$, the critical $K$ when $\tau=10^4P_1$ as estimated by equation~\eqref{eq:Ktau-fit}; (iii) $K_{\tau=1e2}$, the largest $K$ that allows instability when $\tau=10^2P_1$; (iv) $\widetilde{K}_{\tau=1e2}$, the critical $K$ when $\tau=10^2P_1$ as estimated by equation~\eqref{eq:Ktau-fit}. 
    }
    \label{tab:K-fric-2p}
    \begin{tabular}{ccccc} 
    \hline
    $\mu$ & $K_{\tau=1e4}$ & $\widetilde{K}_{\tau=1e4}$ & $K_{\tau=1e2}$ & $\widetilde{K}_{\tau=1e2}$\\
    \hline
    $10^{-6}$ & 3.4 & 3.4 & 2.9 & 2.8 \\
    $10^{-5}$ & 3.3 & 3.3 & 3.1 & 2.9 \\
    $10^{-4}$ & 3.2 & 3.7 & 3.1 & 2.9 \\ 
    $10^{-3}$ & 3.4 & 3.4 & 2.7 & 2.7 \\ 
    \hline
    \end{tabular}
\end{table}

Fig.~\ref{fig:N2-wf-Tinst} compares $t_{\rm inst}$'s from the with-friction runs (coloured dots) to the results from the no-friction runs (black dots). 
Similar to Fig.~\ref{fig:Hill-wRH-wf-fraction}, instability can only happen inside the shaded area with $K \in [2.0,\widetilde{K}_{\tau}]$ and $t_{\rm inst}\in[0,\tau]$ (See Table~\ref{tab:K-fric-2p} for the values of $\widetilde{K}_{\tau}$, as well as the numerical $K_{\tau}$).
Outside of the shaded area, $t_{\rm inst}>\tau$ and instability is suppressed. 
However, because $t_{\rm inst}$ spreads a couple orders of magnitude for runs with the same $\mu$ and $K$, there is no single $K$ value that completely separates stability and instability (see Fig.~\ref{fig:frac-N2}).

\subsection{Three-planet systems}

\begin{figure}
    \includegraphics[width=\columnwidth]{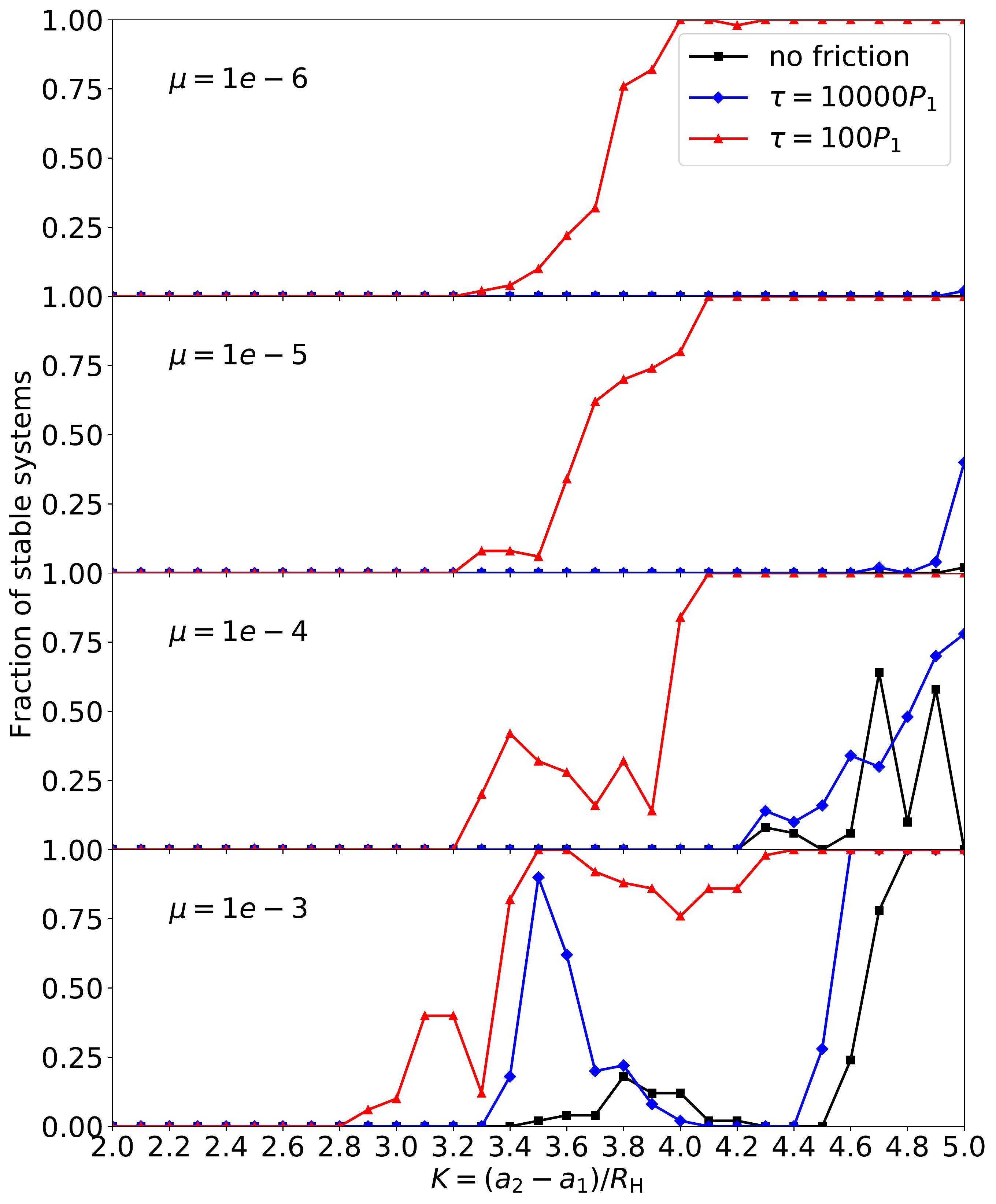}
    \caption{Same as Fig.~\ref{fig:frac-N2}, except showing the fraction of stable three-planet systems after $10^6P_1$. 
    }
    \label{fig:frac-N3}
\end{figure}

Similar to our two-planet simulations, we set up the three-planet simulations with $\mu=10^{-6}$, $10^{-5}$, $10^{-4}$, $10^{-3}$ and repeat each suite three times for ``no-friction'', $\tau=10^2P_1$, and $\tau=10^4P_1$. 
We investigate a wider range of the initial $K$ that includes $2.1$, $2.2$, $...$, $5.0$ because the Hill criterion of stability does not apply in systems with more than two planets. 
Fig.~\ref{fig:frac-N3} summarises the fraction of systems that remain stable after $10^6P_1$ in our simulations.

\begin{figure}
    \includegraphics[width=\columnwidth]{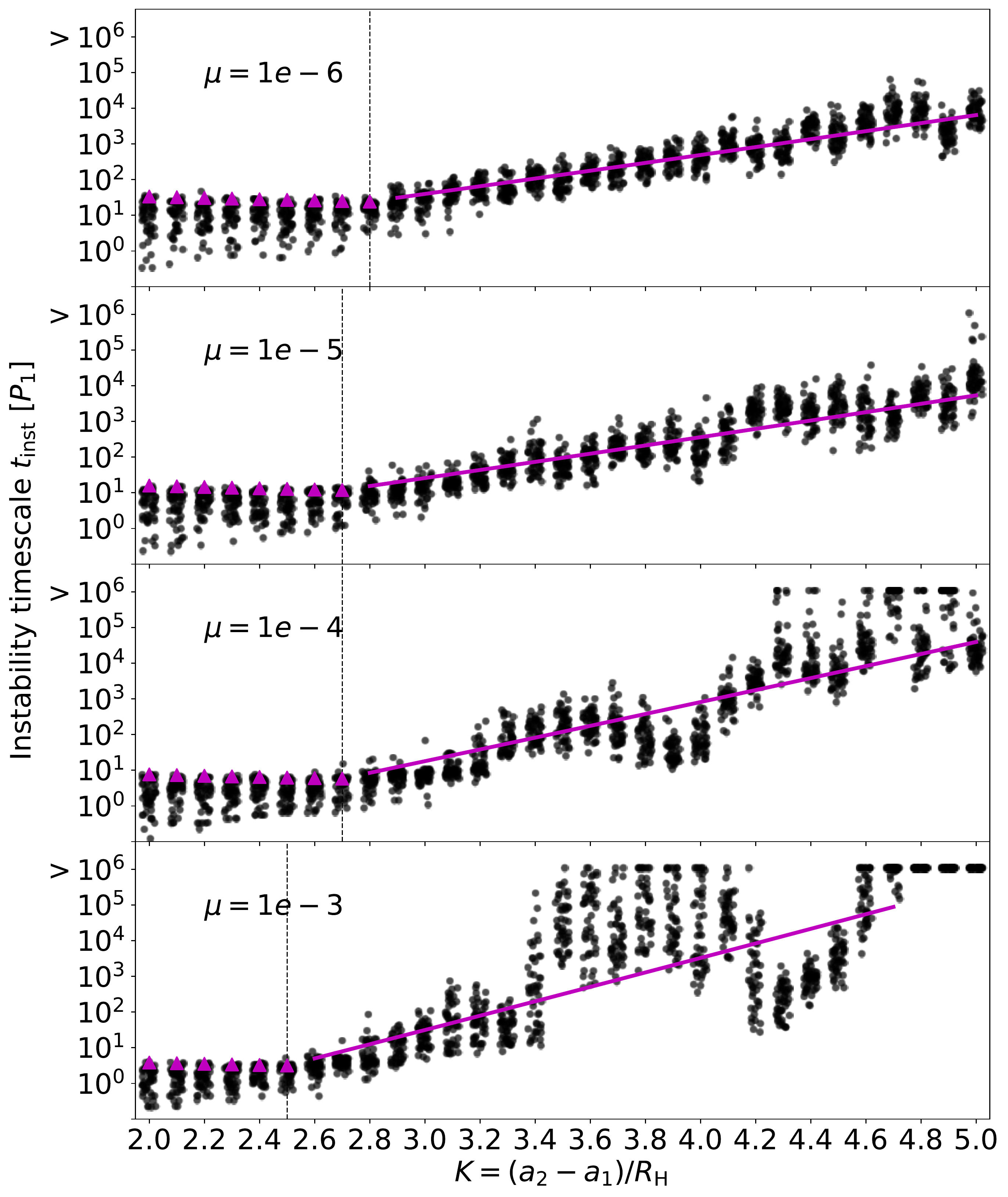}
    \caption{Same as Fig.~\ref{fig:N2-nf-Tinst}, but for three-planet systems. The vertical dashed lines indicate $K_{\rm syn}$.
    }
    \label{fig:N3-nf-Tinst}
\end{figure}

Fig.~\ref{fig:N3-nf-Tinst} shows the instability growth time $t_{\rm inst}$ in the no-friction runs: most of the runs become unstable within $10^{6}P_1$.
The exceptions are at $K\geq4.6$ for $\mu=10^{-3}$, where $t_{\rm inst}$ is too long due to the large initial planet spacings, and similarly at $K=3.8$ for $\mu=10^{-3}$ and $K=4.7$ and $4.9$ for $\mu=10^{-4}$, where $t_{\rm inst}$ are larger than at their neighbouring $K$'s possibly due to mean motion resonances.
We expect the stable fraction (Fig.~\ref{fig:frac-N3}) to converge to zero for all considered $K$ values if we evolve the systems for more orbits.
We find the value of $K_{\rm syn}$ for each $\mu$ using the synodic period of the two inner planets and fit the data points with $K>K_{\rm syn}$ and $t_{\rm inst}<10^6P_1$ using equation~\eqref{eq:Hill-wRH-Tinst-fit}. 
Table~\ref{tab:Tinst-fit-3P} lists the results.

\begin{table}
	\centering
    \caption{Parameters for the instability timescale in the three-planet systems (of different mass ratio $\mu=m_2/M$) with no frictional force. (i) $K_{\rm syn}$, the largest $K$ value that gives $t_{\rm inst}<T_{\rm syn}$ for 80\% of the runs; (ii) $b$, the best fit parameter $b$ in equation~\eqref{eq:Hill-wRH-Tinst-fit} with $1\sigma$ error. 
    }
    \label{tab:Tinst-fit-3P}
    \begin{tabular}{ ccc } 
    \hline
     $\mu$ & $K_{\rm syn}$ & $b$\\
    \hline
     $10^{-6}$ & 2.8 &  $7.8\pm0.1$\\
     $10^{-5}$ & 2.7 &  $7.9\pm0.1$\\
     $10^{-4}$ & 2.7 &  $11.0\pm0.1$\\ 
     $10^{-3}$ & 2.5 &  $12.2\pm0.2$\\ 
    \hline
    \end{tabular}
\end{table}

When the frictional force with $\tau=10^4P_1$ is applied, the survival rate increases at large $K$.
Notably, around half of the systems with $\mu=10^{-5}$ are stable at $K=5.0$. 
Systems with $\mu=10^{-4}$ have a $75\%$ chance to be stable at $K=5.0$ and more than $50\%$ chance to survive at $K\geq 4.8$.
In the $\mu=10^{-3}$ case, $100\%$ of our simulations with $K\geq4.6$ are stable. 
The stable fraction increases dramatically from $\sim0$ to $90\%$ at $K=3.5$.

When $\tau=10^{2}P_1$, we find that stability is guaranteed for $K>4.2$, $4.0$, $4.0$ and $4.3$ for $\mu=10^{-6}$, $10^{-5}$, $10^{-4}$, and $10^{-3}$, respectively. The minimum $K$ for the survival rate to be non-zero is $2.9$ for $\mu=10^{3}$ and $3.3$ for the other three cases.

\begin{figure}
    \includegraphics[width=\columnwidth]{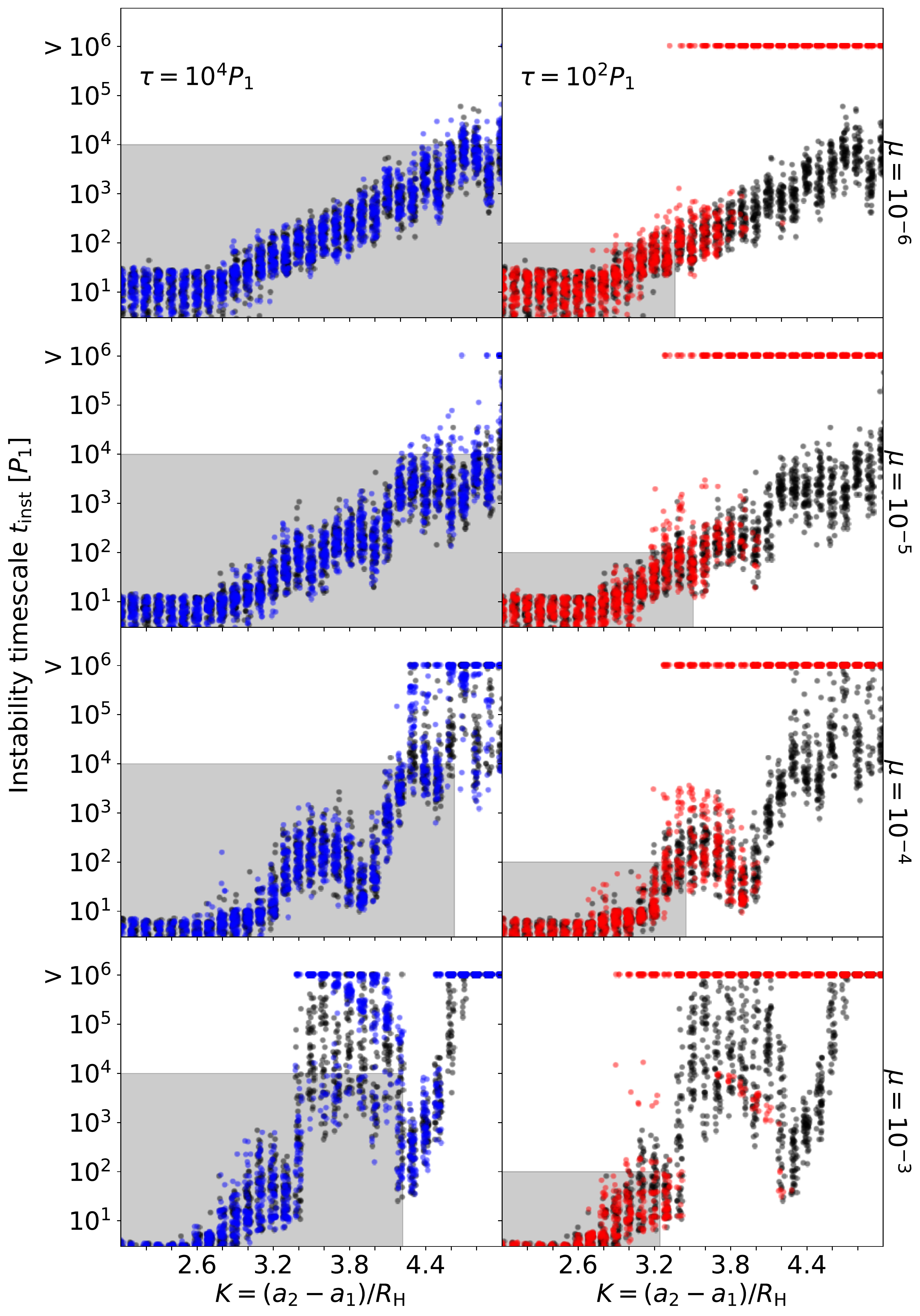}
    \caption{Same as Fig.~\ref{fig:N2-wf-Tinst}, but for three-planet systems.
    }
    \label{fig:N3-wf-Tinst}
\end{figure}

Fig.~\ref{fig:N3-wf-Tinst} displays $t_{\rm inst}$ when the frictional force is applied. 
Unlike in Figs.~\ref{fig:Hill-wRH-wf-Tinst} and~\ref{fig:N2-wf-Tinst}, we see a significant amount of coloured dots (from the with-friction runs) outside of the shaded area (i.e. $[2.0,\widetilde{K}_{\tau}]\times[0,\tau]$).
The stability in three-planet systems cannot be simply determined by the condition $t_{\rm inst}>\tau$ like in the two-planet case.
Nevertheless, $t_{\rm inst}$ still provides an estimate on what value $\tau$ is needed to stabilise a three-planet system.
Our results suggest that, a planetary system may not be stabilised by the frictional force even if $t_{\rm inst}\simeq10-100\tau$. 
Therefore, the critical initial $K$ for stabilisation is larger than $\widetilde{K}_{\tau}$ that one would estimate using equation~\ref{eq:Ktau-fit} (see Table~\ref{tab:K-fric-3p}).

\begin{table}
	\centering
    \caption{Critical initial $K$ when the frictional force (equation~\ref{eq:wf-drag}) is applied in the three-planet systems for different the mass ratio $\mu=m_2/M$. (i) $K_{\tau=1e4}$, the largest $K$ value that allows instability when $\tau=10^4P_1$; (ii) $\widetilde{K}_{\tau=1e4}$, the critical $K$ when $\tau=10^4P_1$ as estimated by equation~\eqref{eq:Ktau-fit}; (iii) $K_{\tau=1e2}$, the largest $K$ that allows instability when $\tau=10^2P_1$; (iv) $\widetilde{K}_{\tau=1e2}$, the critical $K$ when $\tau=10^2P_1$ as estimated by equation~\eqref{eq:Ktau-fit}. 
    }
    \label{tab:K-fric-3p}
    \begin{tabular}{ccccc} 
    \hline
    $\mu$ & $K_{\tau=1e4}$ & $\widetilde{K}_{\tau=1e4}$ & $K_{\tau=1e2}$ & $\widetilde{K}_{\tau=1e2}$\\
    \hline
    $10^{-6}$ & $>5.0$ & $>5.0$ & 4.2 & 3.4 \\
    $10^{-5}$ & $>5.0$ & $>5.0$ & 4.0 & 3.5 \\
    $10^{-4}$ & $>5.0$ & 4.6 & 4.0 & 3.4 \\ 
    $10^{-3}$ & 4.5 & 4.2 & 4.3 & 3.2 \\ 
    \hline
    \end{tabular}
\end{table}

\section{Conclusion}
\label{sec:summary}

\subsection{Summary of key results}

In this paper, we have studied the dynamical instability of closely-packed multi-planet systems in the presence of frictional forces that arise from planet-disc interactions. 
The systems have initially co-planar, near circular orbits, with the orbital separation characterised by the dimensionless ratio $K=(a_{j+1}-a_{j})/R_{\rm H}$ (where $a_j$ is the semi-major axis of the $j$-th planet, and $R_{\rm H}$ is the mutual Hill radius). 
Instability occurs due to planetary orbital crossing (see equation~\ref{eq:Hill-wRH-stop-condition}) as the system evolves.  
The goal of this paper is to evaluate how frictional forces (of various strengths) affect the growth of instability as a function of $K$ and the planet-to-star mass ratio $\mu$.  Note that although we use the terms ``planet'' and ``star'' throughout this paper, our results are also relevant for understanding the evolution of stellar-mass black holes embedded in AGN discs around supermassive black holes.

We consider both ``2 planets'' and ``3 planets'' systems using numerical $N$-body integrations, and carry out theoretical analysis of the restricted three-body problem (star, planet and test particle) to gain analytical understanding. For each planetary architecture, we adopt a range of planet-to-star mass ratios ($\mu$), initial dimensionless orbital separations ($K$) and frictional damping timescales ($\tau$), and determine the fraction of stable systems and instability growth time $t_{\rm inst}$ (i.e., the time to reach the first orbital crossing from initially circular orbits). 
In general, we find that the stable (large $t_{\rm inst}$ and $K$) and unstable (small $t_{\rm inst}$ and $K$) regimes are separated by a grey/transition zone, which can have complicated stable fraction-vs-$K$ dependence because of mean-motion resonances; this transition zone becomes ``smoother'' for more extreme mass ratios ($\mu\lesssim 10^{-6}$). 
Frictional forces tend stabilise a system, pushing the unstable regime towards smaller $K$'s.

For systems with no frictional forces, our key results can be summarised as follows:
\begin{itemize}
\item In the ``planet + test-mass'' and ``2 planets'' simulations, the fraction of stable systems is $100\%$ for $K > K_{\rm crit} \simeq 2\sqrt{3}$ and is less than $1\%$ for $K < K_{\rm gz}$, where $K_{\rm gz}\in[2.6,3.1]$ depending on the mass ratio $\mu$ (see Tables~\ref{tab:Tinst-fit-RH} and~\ref{tab:Tinst-fit-2P}). 
The region $K \in [K_{\rm gz}, K_{\rm crit}]$ is a grey zone where there is no guarantee whether dynamical instability will occur or not (see Figs.~\ref{fig:Hill-wRH-nf-fraction} and~\ref{fig:frac-N2}).
Inside the grey zone, the stable fraction does not always increase monotonically with $K$ because of mean-motion resonances.
In the test-mass cases with $\mu\geq 10^{-5}$, we find stable islands at some particular $K$'s where the stable fraction is enhanced (see Fig.~\ref{fig:Hill-wRH-nf-fraction}).
\item The instability timescales $t_{\rm inst}$ for different planetary architectures are shown in Figs.~\ref{fig:Hill-wRH-nf-Tinst},~\ref{fig:N2-nf-Tinst} and \ref{fig:N3-nf-Tinst}.  
For systems with the same $\mu$ and initial $K$, $t_{\rm inst}$ can vary by 2 to 3 orders of magnitude due its dependence on the initial orbital phases of the planets. In the transition zone, the averaged $t_{\rm inst}$ generally has a non-monotonic dependence on $K$.
\item When $K\leq K_{\rm syn}<K_{\rm gz}$ (see Tables~\ref{tab:Tinst-fit-RH} and~\ref{tab:Tinst-fit-2P}), most simulations exhibit $t_{\rm inst}<T_{\rm syn}$ (the synodic period), i.e., the instability occurs at the first orbital conjunction. When $K > K_{\rm syn}$, the average $t_{\rm inst}$ approximately follows equation (\ref{eq:Hill-wRH-Tinst-fit}).
\end{itemize}

We then re-run the simulations with a frictional force that damps the planetary eccentricity (see equation~\ref{eq:wf-drag}). There are two main effects associated with the frictional force:
\begin{itemize}
\item The frictional force generally increases the stable fraction in all simulations.  
To be stabilised, a two-planet system needs to have $t_{\rm inst}\gtrsim\tau$ (see Figs.~\ref{fig:Hill-wRH-wf-Tinst} and~\ref{fig:N2-wf-Tinst}) and a three-planet system needs to have $t_{\rm inst}\gtrsim10\tau-100\tau$ (see Fig.~\ref{fig:N3-wf-Tinst}).
\item In a two-planet system, the frictional force may lower the planet survival fraction in the stable islands.  
A weak friction force can smooth the stable fraction vs $K$ curve inside the transition zone.  
When there are three planets, the frictional force can make the systems with relatively large $K$ to achieve long-term stability.  
Due to the spread of $t_{\rm inst}$ at the same $K$, a three-planet system may acquire ``grey zone'' when the frictional force is applied (see Fig.~\ref{fig:frac-N3}).
\end{itemize}

We have also devised a linear map to analyse the dynamical instability of the ``planet + test-mass'' system (see Section~\ref{sec:map}).
This map serves as a useful model that captures the key features of the stability, especially the effect of the damping force.

\subsection{Discussion}

Our results are useful for understanding the evolution of multi-planet systems born in protoplanetary discs, and the evolution of multiple stellar-mass black holes (BHs) in ANG discs (see Section~\ref{sec:intro}).
Multiple planets can be brought into closely-packed orbits due to differential migration or migration traps in the disc.  
The frictional forces from the disc acting on the planets can help maintain the stability of the system even for small planetary spacings -- ``how small'' depends on the frictional damping timescale, which in term depends on the disc density.

In this paper, we have focused on the onset of dynamical instability, signalled by close encounters between two planets (see equation.~\ref{eq:Hill-wRH-stop-condition}). 
Once the instability is initiated, the system will experience chaotic orbital evolution.  
In the absence of gas friction, close encounters keep recurring until one planet is ejected, or two planets collide with each other, or one planet crashes into the star.
The branching ratio of each outcome depends on the planetary mass, radius and orbital velocity around the host star \citep[see][and references therein]{Li2021}. 
In the case of stellar-mass BHs around a supermassive BH, the large mass ratio implies that ejection is highly unlikely or impossible; and the chaotic orbital evolution is terminated when two BHs undergo an extremely close encounter and form a bound, merging binary due to gravitational radiation \citep[see][and references therein]{Li2022}.

When dynamical instability develops in the presence of gas discs, there exists another possible outcome in which the system regains its stability after the chaotic evolution:  the combined effects of close encounters and frictional forces may push the planets (or BHs) into well separated orbits to ensure dynamical stability with friction \citep{Li2022}.
In particular, we expect that some planetary ejections and star crashings may be prevented by the frictional forces because they typically require a large number of closer encounters than planet-planet collisions.

\section*{Acknowledgements}
This work is supported in by NSF grant AST-2107796 and the NASA grant 80NSSC19K0444.
Jiaru Li is supported by the Center for Space and Earth Science at Los Alamos National Laboratory (approved for public release as LA-UR-22-23672).
This research used resources provided by the Los Alamos National Laboratory Institutional Computing Program, which is supported by the U.S. Department of Energy National Nuclear Security Administration under Contract No. 89233218CNA000001.

\section*{Data Availability}

The data underlying this article will be shared on reasonable request to the corresponding author.



\bibliographystyle{mnras}



\appendix

\section{Derivation of the Mean-Motion Resonance overlap scaling from the linear map}
\label{sec:map-appendix}

This derivation has been proposed by DQT89. Let us consider the first conjunction of the linear map presented in Section~\ref{sec:map}. DQT89 argues that the onset of chaos occurs when the variable part of the conjunction longitude difference $|\lambda_n-\lambda_{n-}|$ is similar or greater than $\pi$. From equation~\eqref{eq:map-delta-lambda}, we have
\begin{equation}
    |\lambda_n-\lambda_{n-}| \simeq \frac{4\pi}{3\varepsilon_1}
    \simeq \frac{4\pi}{3\varepsilon_{1-}} \left(1-\frac{\Delta\varepsilon}{\epsilon_{1-}}\right),
\end{equation}
where $\Delta\varepsilon = \epsilon_{1} - \epsilon_{1-}$. DQT89's condition thus writes:
\begin{equation}
    \frac{4\pi}{3} \frac{\Delta\varepsilon}{\varepsilon_{1-}^2} \gtrsim \pi.
\end{equation}
We assume that the orbits are initially circular, so that $\Delta\varepsilon \simeq 2|z_1|^2 / (3 \varepsilon_{1-})$ (equation~\ref{eq:map-eps-step}) and $|z_1| \simeq g\mu/\varepsilon_{1-}^2$ (equation~\ref{eq:map-z-step}). It then leads to the condition
\begin{align}
   &\frac{4\pi}{3} \frac{2}{3\varepsilon_{1-}^3} \frac{g^2\mu^2}{\varepsilon_{1-}^4} \gtrsim \pi,\\
   &\iff \varepsilon_{1-} \lesssim \left(\frac{8}{9}\right)^\frac{1}{7} g^\frac{2}{7} \mu^\frac{2}{7}.
\end{align}


\bsp	
\label{lastpage}
\end{document}